\documentclass[fleqn,usenatbib]{mnras}
\usepackage{newtxtext,newtxmath}

\usepackage[T1]{fontenc}
\usepackage{ae,aecompl}
\usepackage{times}
\usepackage{rotating}
\usepackage{graphicx}	
\usepackage{amsmath,amstext}	
\usepackage{amssymb}	
\usepackage[all]{hypcap} 
\usepackage{bm}
\usepackage[usenames]{xcolor}

\newcommand{\Msun}{\,M_{\odot}}

\newcommand{\epsint}{\epsilon_{\mathrm{int}}}
\newcommand{\tff}{\tau_{\mathrm{ff}}}
\newcommand{\tdur}{\tau_{\mathrm{dur}}}
\newcommand{\tspread}{\tau_{\mathrm{spread}}}
\newcommand{\fb}{f_{\mathrm{boost}}}
\newcommand{\fbd}{f_{\mathrm{bound}}}
\newcommand{\Mgmc}{M_{\mathrm{GMC}}}
\newcommand{\Rgmc}{R_{\mathrm{GMC}}}
\newcommand{\vrms}{v_{\mathrm{rms}}}

\title[Bound star clusters emerged from GMCs]{Disruption of giant molecular clouds and formation of bound star clusters under the influence of momentum stellar feedback}

\author[H. Li et al.]{
Hui Li$^{1}$\thanks{E-mail: hliastro@mit.edu}\href{https://orcid.org/0000-0002-1253-2763}{\includegraphics[scale=0.6]{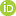}},
Mark Vogelsberger$^{1}$,
Federico Marinacci$^{3,2,1}$\href{https://orcid.org/0000-0003-3816-7028}{\includegraphics[scale=0.6]{orcid.png}}
Oleg Y. Gnedin$^{4}$\href{https://orcid.org/0000-0001-9852-9954}{\includegraphics[scale=0.6]{orcid.png}}
\\
$^{1}$Department of Physics, Kavli Institute for Astrophysics and Space Research, Massachusetts Institute of Technology, Cambridge, MA 02139, USA\\
$^{2}$Harvard-Smithsonian Center for Astrophysics, 60 Garden Street, Cambridge, MA 02138, USA\\
$^{3}$Department of Physics \& Astronomy, University of Bologna, via Gobetti 93/2, 40129 Bologna, Italy\\
$^{4}$Department of Astronomy, University of Michigan, Ann Arbor, MI 48109, USA
}

\date{Accepted XXX. Received YYY; in original form ZZZ}

\pubyear{2018}

\begin{document}
\label{firstpage}
\pagerange{\pageref{firstpage}--\pageref{lastpage}}
\maketitle

\begin{abstract}
Energetic feedback from star clusters plays a pivotal role in shaping the dynamical evolution of giant molecular clouds (GMCs). To study the effects of stellar feedback on the star formation efficiency of the clouds and the dynamical response of embedded star clusters, we perform a suite of isolated GMC simulations with star formation and momentum feedback subgrid models using the moving-mesh hydrodynamics code \textsc{Arepo}. The properties of our simulated GMCs span a wide range of initial mass, radius, and velocity configurations.
We find that the ratio of the final stellar mass to the total cloud mass, $\epsint$, scales strongly with the initial cloud surface density and momentum feedback strength. This correlation is explained by an analytic model that considers force balancing between gravity and momentum feedback. For all simulated GMCs, the stellar density profiles are systematically steeper than that of the gas at the epochs of the peaks of star formation, suggesting a centrally concentrated stellar distribution. We also find that star clusters are always in a sub-virial state with a virial parameter $\sim0.6$ prior to gas expulsion. Both the sub-virial dynamical state and steeper stellar density profiles prevent clusters from dispersal during the gas removal phase of their evolution. The final cluster bound fraction is a continuously increasing function of $\epsint$. GMCs with star formation efficiency smaller than 0.5 are still able to form clusters with large bound fractions.
\end{abstract}

\begin{keywords}
methods: numerical -- stars: formation -- stars: kinematics and dynamics -- galaxies: star clusters: general
\end{keywords}



\section{Introduction}\label{sec:intro}
Most, if not all, stars are formed in clusters \citep{lada_lada03}, which emerged from giant molecular clouds \citep[GMC;][]{shu_etal87, scoville_good89,mckee_ostriker07, krumholz_etal18}. Due to the complex interplay of gravity, supersonic turbulence and stellar feedback from massive stars, the dynamical evolution and cluster formation activities in GMCs are still highly debatable \citep[e.g.][]{krumholz_mckee05,ballesteros-paredes_hartmann07,heitsch_etal09,hennebelle_chabrier11, padoan_nordlund11,federrath_klessen12,burkert_hartmann13,traficante_etal18}. One of the key
observables that can be used to constrain various physical process is the star formation efficiency (SFE) in star-forming regions.

It is well known that star formation is inefficient on galactic scales. The observed linear correlation between molecular gas surface density and star formation rate (SFR) surface density in normal star-forming galaxies suggests a nearly constant gas depletion time-scale around $\sim$~2~Gyr, much longer than the dynamical time-scale of galactic disks \citep{kennicutt89,bigiel_etal08,saintonge_etal11, leroy_etal13,genzel_etal15,tacconi_etal18}. In contrast, the SFE on GMC scales shows a large
variation ranging from less than a few percent to nearly unity \citep{zuckerman_evans74,krumholz_tan07, wu_etal10,evans_etal14,heyer_etal16,lee_etal16,vutisalchavakul_etal16}. The origin of this large scatter is usually explained as a combination of the time variability of the SFR during the course of cloud evolution and intrinsic scatter of SFEs due to the diversity of GMC properties \citep{feldmann_gnedin11, kruijssen_longmore14,lee_etal16,grudic_etal18b,kruijssen_etal18b}.
For example, recent theoretical models and high-resolution magneto-hydrodynamics simulations suggest that the SFE depends on the local virial parameters of the cloud controlled by large-scale turbulence \citep[e.g.][]{krumholz_mckee05, padoan_etal12}. However, it has recently been recognized that large-scale turbulence can only account for an $\sim$~0.3~dex scatter, which is not sufficient to explain the observed SFE variations \citep{lee_etal16}. Another source of variation comes from different stellar feedback channels that alter the dynamical states of the GMCs \citep{fall_etal10,murray_etal10,dale_etal14,myers_etal14,raskutti_etal16,kim_etal17,grudic_etal18}.
Previous studies found that GMC simulations adopting different stellar feedback mechanisms (stellar winds, ionizing radiation, or supernovae) lead to dramatically different final SFEs. The problem has recently been recognized to be more subtle than previously thought, since even small differences in numerical treatments, such as different radiative transfer schemes, massive star sampling, and momentum and energy deposition algorithms, can lead to drastic changes for the final
SFE \citep{dale_etal05,roskar_etal14,raskutti_etal16,grudic_etal18,kim_etal18}. Therefore, how the SFE depends on GMC properties and the strength of stellar feedback remains an open question.

Stellar feedback not only changes the efficiency of star formation within GMCs, but also alters the dynamical state of star clusters by dispersing the cloud.
The process of star cluster disruption due to rapid gas expulsion shortly after the cluster emerges from its natal cloud is believed to be the main culprit of the ``infant mortality'' of star clusters -- a sharp decrease in the number of young star clusters with the increase of cluster age in local star-forming regions \citep[e.g.][]{lada_lada03}.
A simple virial analysis suggests that a star cluster that is initially in virial equilibrium will dissociate if more than half of the mass is instantaneously lost \citep{hills80,mathieu83}. However, this statement does not take into account the highly non-linear star formation and stellar feedback process in realistic self-gravitating turbulent environments. 
For example, embedded clusters in star-forming regions are not necessarily in virial equilibrium. Recent hydrodynamical simulations suggest that the stellar velocity dispersions are in general much smaller than that of the gas, suggesting a sub-virial dynamical state of star clusters within GMCs \citep[e.g.][]{offner_etal09}.
Moreover, stars are usually not well mixed with gas but instead formed in the densest part of the cloud. The difference between the gas and stellar distribution can strongly affect the dynamical response of star clusters to gas dispersal \citep[e.g.][]{kruijssen_etal12b,shukirgaliyev_etal18}.
Most importantly, GMCs are highly substructured. Stars are formed at the intersections of gas filaments and assembled into different subclusters hierarchically.
Previous works have explored some of the above complications using different physical and numerical methods, including analytical models \citep{hills80, mathieu83, adams00, boily_kroupa03a, kruijssen12, parmentier_pfalzner13}, pure \textit{N}-body simulations \citep{tutukov78, lada_etal84,boily_kroupa03b, goodwin_bastian06,baumgardt_kroupa07, smith_etal11,farias_etal18}, and hydrodynamic simulations \citep{bonnell_etal11, girichidis_etal12,moeckel_etal12,fujii_portegies_zwart_2016,gavagnin_etal17}. 
Recent efforts have been made to include various relevant physical processes in hydrodynamic simulations \citep[e.g.][]{parker_etal15, gavagnin_etal17}, however, due to the high computational costs, they usually focus on a handful of less massive GMCs.

In this paper, we perform a suite of hydrodynamic simulations of turbulent GMCs employed with a simple star formation and stellar feedback models in the moving-mesh code \textsc{AREPO}. We survey GMCs with a broad range of mass, size, and velocity configurations to investigate the physical origin of the intrinsic variations of SFEs and the properties of surviving star clusters. The structure of this paper is as follows. In Section~\ref{sec:methods}, we describe the simulation setup, star formation and momentum stellar feedback
implementations, and the design of initial conditions. In Section~\ref{sec:sfe}, we examine the dependence of integrated SFE of GMCs on cloud mass, size, and momentum feedback intensity. In Section~\ref{sec:fbound}, we describe the subsequent dynamical evolution of star clusters after the residual gas is completely expelled and investigate the relationship between SFE and cluster bound fraction.
We summarize our conclusions in Section~\ref{sec:summary}.

\section{Methods}\label{sec:methods}

\subsection{Simulation setup}\label{sec:method-setup}

The simulations in this work are performed with \textsc{Arepo} \citep{springel10arepo}, a moving-mesh, finite-volume hydrodynamic code employing a second-order unsplit Godunov scheme. The control volumes are discretized by a Voronoi tessellation, which is generated from its dual Delaunay tessellation determined by a set of mesh-generating points. These points can move freely within the simulation domain and follow gas flows in a quasi-Lagrangian fashion. Therefore, \textsc{Arepo} captures the advantages of both grid- and particle-based hydrodynamic methods and has already been applied to various astrophysical problems \citep[e.g.][]{keres_etal12, torrey_etal12, vogelsberger_etal12, vogelsberger_etal14, springel_etal18}.
Our simulations include hydrodynamics, self-gravity, radiative cooling, star formation, and momentum feedback from stellar winds.

We use an adaptive softening scheme for gas cells so that the gravitational forces are resolved all the way down to the size of each cell.
We employ a quasi-Lagrangian refinement scheme that keeps the mass of gas cells close to a target mass determined by initial conditions. In addition, we refine a cell if its volume is more than 32 times larger than the minimum volume of all its face-touching neighbours. This volume-limited refinement scheme prevents large volume contrast between adjacent cells and helps to better resolve the regions that experience fast expansion due to stellar feedback.

Since radiative cooling is responsible for gas fragmentation and subsequent star formation in GMCs, it is important to follow the cooling process explicitly over a large range of temperatures. Instead of adopting isothermal or effective equations of state, which have been used in many previous GMC simulations \citep[e.g.][]{dale_etal05,raskutti_etal16,kim_etal18}, 
we explicitly include radiative cooling from multiple channels: a network implementing hydrogen and helium cooling and heating processes due to collisions, recombinations, free-free emission and photoionization from UV background radiation; high-temperature ($T > 10^{4}\, {\rm K}$) metal-line cooling that is added to the hydrogen and helium network following \citet{vogelsberger_etal13}; low-temperature ($T < 10^{4}\, {\rm K}$), metal-line, fine-structure and molecular cooling implemented as a fitting function, depending on temperature, density and gas metallicity, to \textsc{CLOUDY} calculations presented in \citet{hopkins_etal18} and Marinacci et al. (in prep.). We set the metallicity of the GMCs to solar abundance when evaluating the cooling rates from metals.
One caveat is that the adopted cooling curves are based on the \textsc{CLOUDY} model calculations under the spatially uniform UV background used for galaxy formation simulations. The cooling rate calculated based on this uniform UV background may not be accurate for GMC simulations, especially in close proximity of massive stars.

\subsection{Star formation}\label{sec:method-sf}
During each simulation time-step, we identify all star-forming cells and convert them to stellar particles probabilistically.
Star-forming cells are defined as gas cells that are cold ($T_{\rm cell}<100$~K), contracting ($\nabla\cdot \bm{v}<0$), and self-gravitating ($|\nabla\cdot\bm{v}|^2+|\nabla\times\bm{v}|^2<2G\rho$), where $\bm{v}$ and $\rho$ are the velocity and density of the cells, respectively. We also employ a density threshold for star formation, $n_{\rm cell} > 10^5{\rm cm^{-3}}$, to avoid rare situations where some self-gravitating clumps are formed in the very low density outskirt of the cloud. 

A given star-forming cell is converted to stellar particles with a constant probability $p=\Delta t/\tff(\rho)$ at a given time-step $\Delta t$, where $\tff=(3\pi/32G\rho)^{1/2}$ is the free-fall time of the cell. The cells that are converted to stellar particles are removed and the volume of these cells is claimed by their neighbours.
The mass, position, and velocity of the newly formed stellar particles are inherited from their parent gas cells. Therefore, the mass distribution of stars is similar to that of the gas particles, which is around the target mass of the simulations. After the stellar particles are created, they are treated as collisionless particles with a Plummer-equivalent softening length fixed to $10^{-4}$ of the initial diameter of the GMC.

\subsection{Momentum stellar feedback}\label{sec:method-feedback}
The overall evolution of GMCs depends strongly on the strength of stellar feedback. Unfortunately, the exact amount of feedback that is associated with massive stars in the simulations is still debated. As has already been noticed in previous studies, GMC simulations with different stellar feedback sources (stellar winds, ionizing radiation, or supernovae) show dramatically different gas evolution and star formation efficiencies \citep{dale_etal05,roskar_etal14,raskutti_etal16,grudic_etal18c,kim_etal18}. Moreover, it has recently been recognized that, even some small changes in numerical implementations, such as radiation hydrodynamic methods \citep{kim_etal18}, sampling of massive star formation \citep{grudic_etal18b}, and momentum/energy deposition algorithms \citep{hopkins_etal18}, can contribute noticeable variation to the star formation efficiencies of the clouds.
Since exploring accurate feedback implementation from various sources is not the main focus of this paper, we simply treat stellar feedback by depositing mass and momentum fluxes from stellar particles to their neighbouring gas cells.

We set the fiducial mass-loss and momentum deposition rate to the initial mass function (IMF)-averaged values of stellar winds from a single stellar population with a Kroupa initial mass function \citep{kroupa01}. Following \citet{hopkins_etal18}, the mass-loss rate per unit stellar mass is
\begin{equation}\label{eq:wind-mass-loss}
\frac{\dot{m}_{\rm w}}{{\rm Gyr^{-1}}} = \begin{cases}
4.763(0.01+Z/Z_\odot) & t_6<1\\
4.763(0.01+Z/Z_\odot)t_6^{1.45+0.8\ln{(Z/Z_\odot)}} & 1<t_6<3.5\\
29.4(t_6/3.5)^{-3.25} + 0.0042 & 3.5<t_6<100
\end{cases}
\end{equation}
where $Z$ is the metallicity and $t_6$ is the age of stellar particles in unit of Myr.
The kinetic luminosity of winds per unit stellar mass is
\begin{equation}\label{eq:wind-lum}
l_{\rm w} = \left[\frac{5.94\times10^4}{1+(t_6/2.5)^{1.4}+(t_6/10)^5}+4.83\right]\times10^{12}\dot{m}_{\rm w}~{\rm erg/g},
\end{equation}
for $t_6<100$. Winds from stellar particles older than $t_6>100$ are irrelevant here since the dynamical time-scales of our model GMCs are much shorter than 100~Myr.

The mass-loss and wind momentum are deposited to the gas cells around each stellar particle in the following way: for a given stellar particle of mass $m_*$ at time-step $\Delta t$, the total mass-loss is $\Delta m=\dot{m}_{\rm w}m_*\Delta t$ and wind momentum is $\Delta p=\sqrt{2l_{\rm w}\dot{m}_{\rm w}}\fb m_*\Delta t$, where $\fb$ is a boosting factor to the fiducial wind momentum to mimic the feedback intensity from different feedback sources. The mass and momentum fluxes from a stellar particle are distributed to its nearest 32 neighbouring gas cells in a weighted fashion so that cell $i$ with weight $w_i$ receives mass $\Delta m_i=(w_i/\Sigma_j{w_j})\Delta m$ and momentum $\Delta \bm{p}_i=(w_i/\Sigma_j{w_j})\Delta p\bm{r}_i/|\bm{r}_i|$, where $\bm{r}_i$ is the vector from the position of the stellar particle to the mesh-generating point of cell $i$.
The weight can be chosen to be any physical quantities of the cells, such as volume, mass, or solid angle opened to the stellar particle. To test the robustness of the feedback implementation, we perform a series of numerical tests of wind-blowing bubbles created by a stellar particle with constant mass-loss rate $\dot{m}=10^{-5}\Msun/$yr and wind velocity $v_{\rm w}=500$~km/s (see Appendix~\ref{sec:appendix-wind}). We find that the expansion history and the internal structure of the bubble are consistent with analytical solutions in \citet{weaver_etal77}. We also test the sensitivity of the star formation history of one GMC using different weighting methods (see Appendix~\ref{sec:appendix-wind-gmc}). 
To make consistent investigation across all GMCs, in the rest of the paper, we use solid angle as the weight to deposit mass and momentum.

\begin{table*}
\centering
\caption{Model Parameters}
\begin{tabular}{lccccccccc}
\hline
Name		& $M_{\rm GMC} (\Msun)$ & $R$ (pc) & $\alpha_0^R$ & $\tau_{\rm ff}$ (Myr) & $m_{\rm res}$ ($\Msun$) & $l_{\rm soft}$ ($10^{-3}$pc)\\
\hspace{3mm}(i)&(ii)&(iii)&(iv)&(v)&(vi)&(vii)\\
\hline
RHO5 T/R	& $1.25\times10^4$	& 5		& 0.1/0.9	& 1.6	& $6.0\times10^{-3}$& 1\\
RHO10		& $10^5$			& 10	& 0.1/0.9 	& 1.6	& $4.8\times10^{-2}$& 2\\
RHO20		& $8\times10^5$		& 20	& 0.1/0.9 	& 1.6	& $0.38$ 			& 4\\
RHO40		& $6.4\times10^6$	& 40	& 0.1/0.9	& 1.6	& $3.1$ 			& 8\\
RHO80		& $5.12\times10^7$	& 80	& 0.1/0.9   & 1.6	& $24.4$ 			& 16\\
SIGMA5		& $2.5\times10^4$	& 5 	& 0.1/0.9   & 1.13	& $1.2\times10^{-2}$& 1\\
SIGMA20		& $4\times10^5$		& 20	& 0.1/0.9   & 2.26	& $0.19$			& 4\\
SIGMA40		& $1.6\times10^6$	& 40	& 0.1/0.9   & 3.2	& $0.76$			& 8\\
\hline
\end{tabular}\\
\begin{flushleft}
\textbf{Note.} Column information: (i) model name, (ii) initial GMC mass, (iii) initial GMC radius, (iv) initial virial parameter for the rotational components: $\alpha_0^R=2E_{\rm rot}/|E_G|$, see \autoref{sec:method-IC} in details, (v) initial free-fall time, (vi) target mass for gas cells, (vii) gravitational softening length of stellar particles. \\
\end{flushleft}
\label{tab:run-parameter}
\end{table*}

\subsection{Initial conditions}\label{sec:method-IC}
We set up the initial condition of GMCs as gas spheres of uniform density with initial turbulent velocity fields. The mass, radius, and other physical parameters of the initial conditions are listed in Table~\ref{tab:run-parameter}. We choose the initial mass ($\Mgmc$) and radius ($\Rgmc$) of the GMCs so that all ``RHO'' runs have the same initial volume density ($\rho_0=\Mgmc/(\frac{4}{3}\pi\Rgmc^3)\approx24\Msun{\rm pc^{-3}}$) and all ``SIGMA'' runs have the same initial surface density ($\Sigma_0=\Mgmc/\pi\Rgmc^2\approx318\Msun{\rm pc^{-2}}$). The goal of this experimental design is to determine whether SFE depends on volume density or surface density.

The initial velocity field is initialized as a combination of turbulent motions and rigid rotation along the $z$-axis. We assign the rotational velocity field as
\begin{equation}
v_x^R(x,y,z)=-\Omega_iy; \; v_y^R(x,y,z)=\Omega_ix; \; v_z^R(x,y,z)=0,
\end{equation}
where $v_x^R, v_y^R, v_z^R$ are the three components of the rotation velocities with circular frequency $\Omega_i$. For each run listed in \autoref{tab:run-parameter}, we construct two separate initial conditions: rotation-supported (``R'') and turbulent-supported (``T'') runs. The virial parameter contributed from rotation $\alpha_0^R\equiv2E_{\rm rot}/|E_G|=0.1/0.9$ is used to calculate $\Omega_i$ for the corresponding ``T''/``R'' runs, where $E_{\rm rot}$ and $E_G$ are the rotational  energy and gravitational energy of the cloud.

\begin{figure*}
\includegraphics[width=0.985\columnwidth]{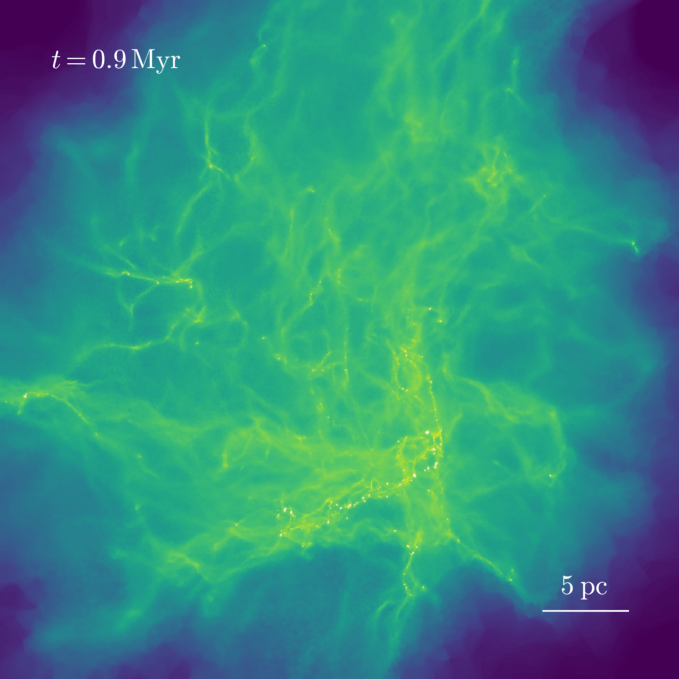}
\hspace{0.5mm}
\includegraphics[width=0.985\columnwidth]{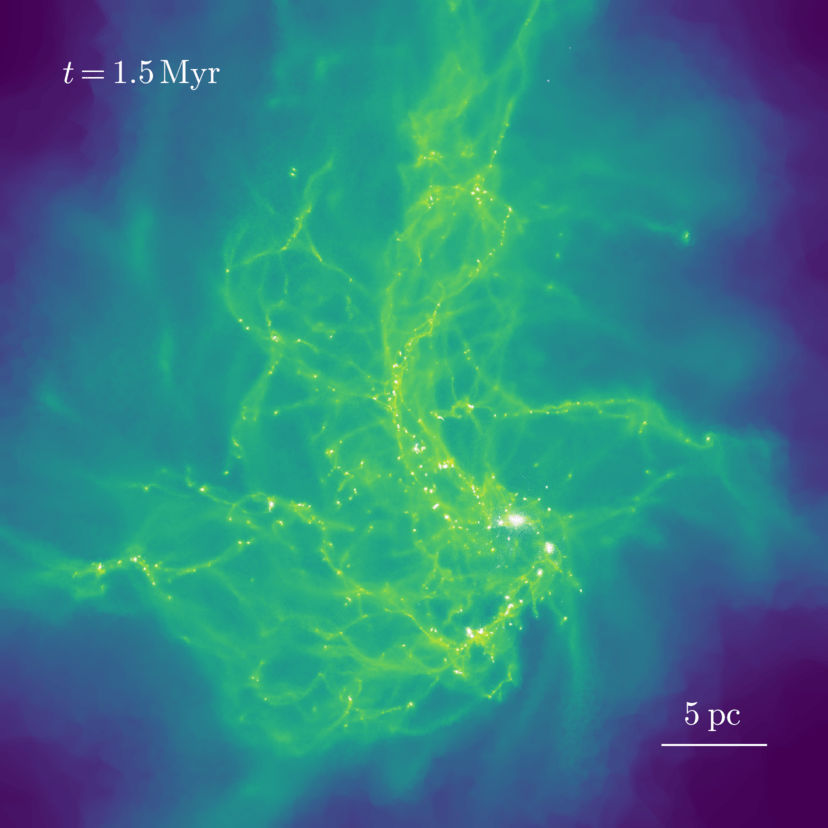}
\vspace{2mm}\vspace{-\lineskip}
\includegraphics[width=0.985\columnwidth]{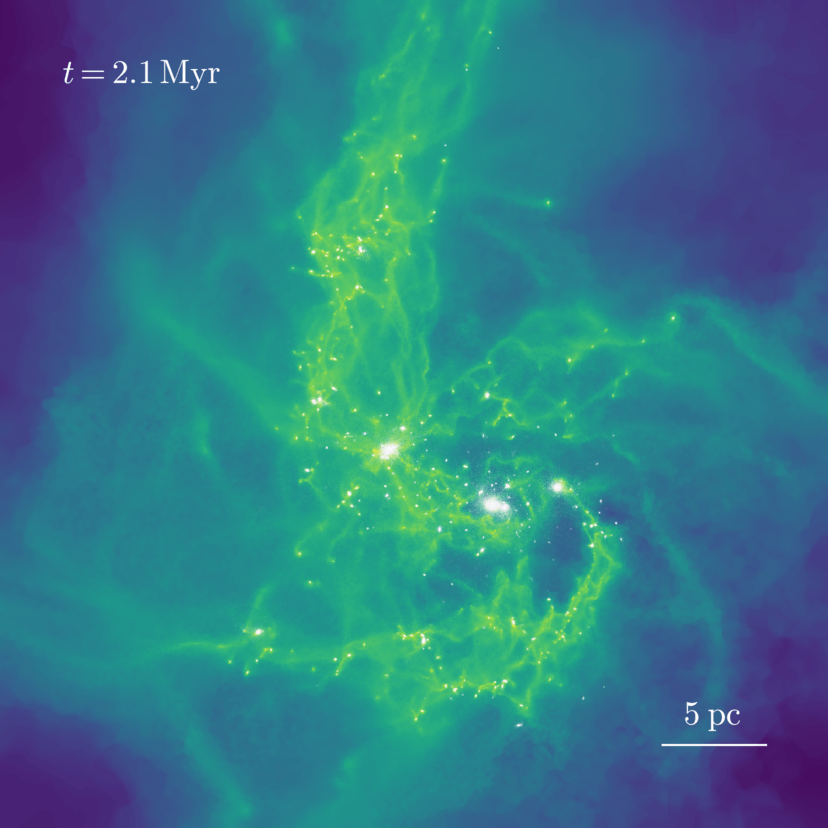}
\hspace{0.5mm}
\includegraphics[width=0.985\columnwidth]{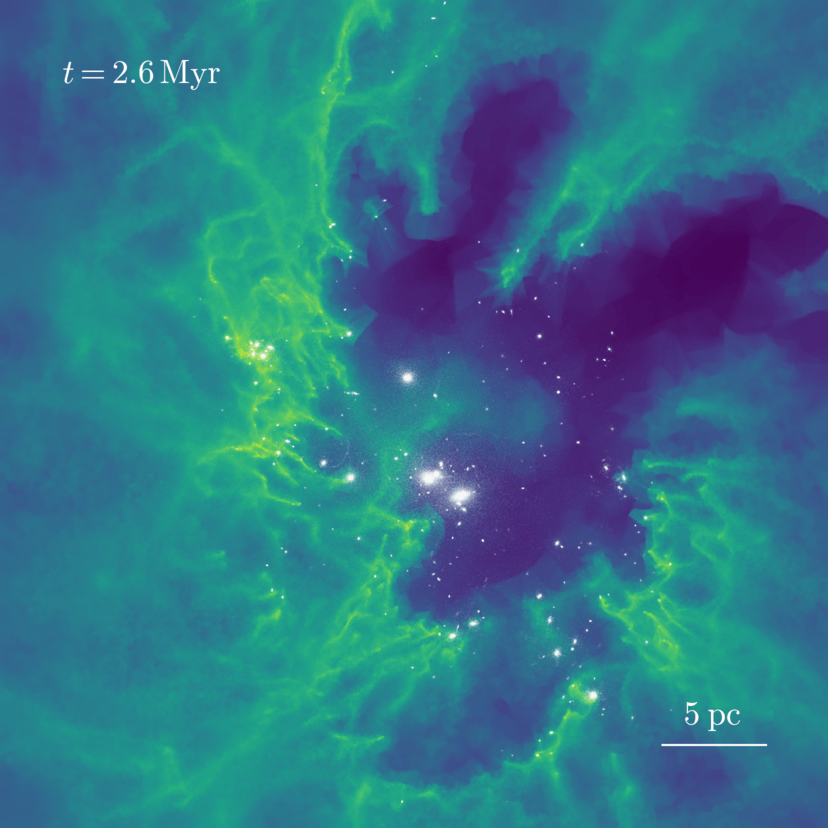}
\vspace{0mm}
\caption{Gas density projection plots for RHO20T run with $\fb=2$ at four epochs: $t=$~0.9, 1.5, 2.1, and 2.6~Myr. The colour range for gas surface density is from $\Sigma_{\rm gas}=50\Msun/{\rm pc}^2$ to $10^6\Msun/{\rm pc}^2$ and a length scale of 5~pc is labelled on the lower right corners in each panel. Stellar particles are presented by white dots.}
  \label{fig:prj_x}
\end{figure*}

The turbulent velocity field is first initialized as a Gaussian random field in the Fourier space with the variance of the field determined by a given power spectrum, $P(k)$. Each dimension of the turbulent velocity field is treated independently and the result is a natural mixture of solenoidal and compressive turbulence. In order to rearrange the turbulent field into arbitrary solenoidal and compressive components, we perform a Helmholtz decomposition in $k$-space by applying the projection operator to the field \citep{federrath_etal10}
\begin{equation}
\bm{\mathcal{H}}_{ij}(\mathbf{k})=\eta_{\rm comp}\bm{\mathcal{H}}_{ij}^{\parallel}(\mathbf{k})+(1-\eta_{\rm comp})\bm{\mathcal{H}}_{ij}^{\bot}(\mathbf{k}),
\end{equation}
where $\bm{\mathcal{H}}_{ij}^{\parallel}(\mathbf{k})=k_ik_j/k^2$ and $\bm{\mathcal{H}}_{ij}^{\bot}(\mathbf{k})=\bm\delta_{ij}-k_ik_j/k^2$ are the compressive and solenoidal operators, respectively, and $\bm\delta_{ij}$ is the Kronecker delta function. $\eta_{\rm comp}$ is the contribution from the compressive mode ranging from 0 to 1. $\eta_{\rm comp}=1$ means a purely compressive turbulence field and $\eta_{\rm comp}=0$ means a purely solenoidal. Varying $\eta_{\rm comp}$ would lead to a different density structure of the clouds. For simplicity, we use $\eta_{\rm comp}=1/3$ so that the ratio of kinetic energy of the compressive mode to that of the solenoid mode is 2:1 to mimic the natural mixture of the two turbulent modes.
After projection, the turbulence field in $k$ space is Fourier-transformed to real space and is then interpolated to the position of gas cells within the sphere. The field is renormalized so that the virial parameter due to turbulence is $\alpha_0^T\equiv2E_{\rm turb}/|E_G|$.
We adopt a power-law power spectrum $P(k)\propto k^{-4}$, which is similar to the turbulence properties of GMCs \citep[e.g.][]{dobbs_etal14}. We assume the cloud is initially in virial equilibrium so that the virial parameter of the cloud $\alpha_0=\alpha_0^R+\alpha_0^T=1$.

The gas temperature is initialized to 10~K, which is commonly used for GMC simulations \citep{dale_etal14,raskutti_etal16}. The choice of the initial temperature does not change the evolution of the gas and star of the GMCs because of the short cooling time-scale compared to the dynamical time-scale of the clouds. For each initial condition, we perform five simulations with different momentum boosting factor, $\fb=$~0.5, 1, 2, 4, 10, to test the effects of the strength of momentum feedback on the global evolution of GMCs and star clusters. GMCs are initially resolved by $128^3$ equal-mass gas cells, which sets the target mass $m_{\rm res}\approx\Mgmc/128^3$. We perform a convergence test for the RHO20T run with $\fb=2$ by varying the number of resolution elements from $64^3$ to $256^3$ and find that the star formation histories are not sensitive to mass resolutions. The final star formation efficiencies in runs with different resolutions vary by only a few percent.

\begin{figure}\label{fig:time-evolution}
\includegraphics[width=\columnwidth]{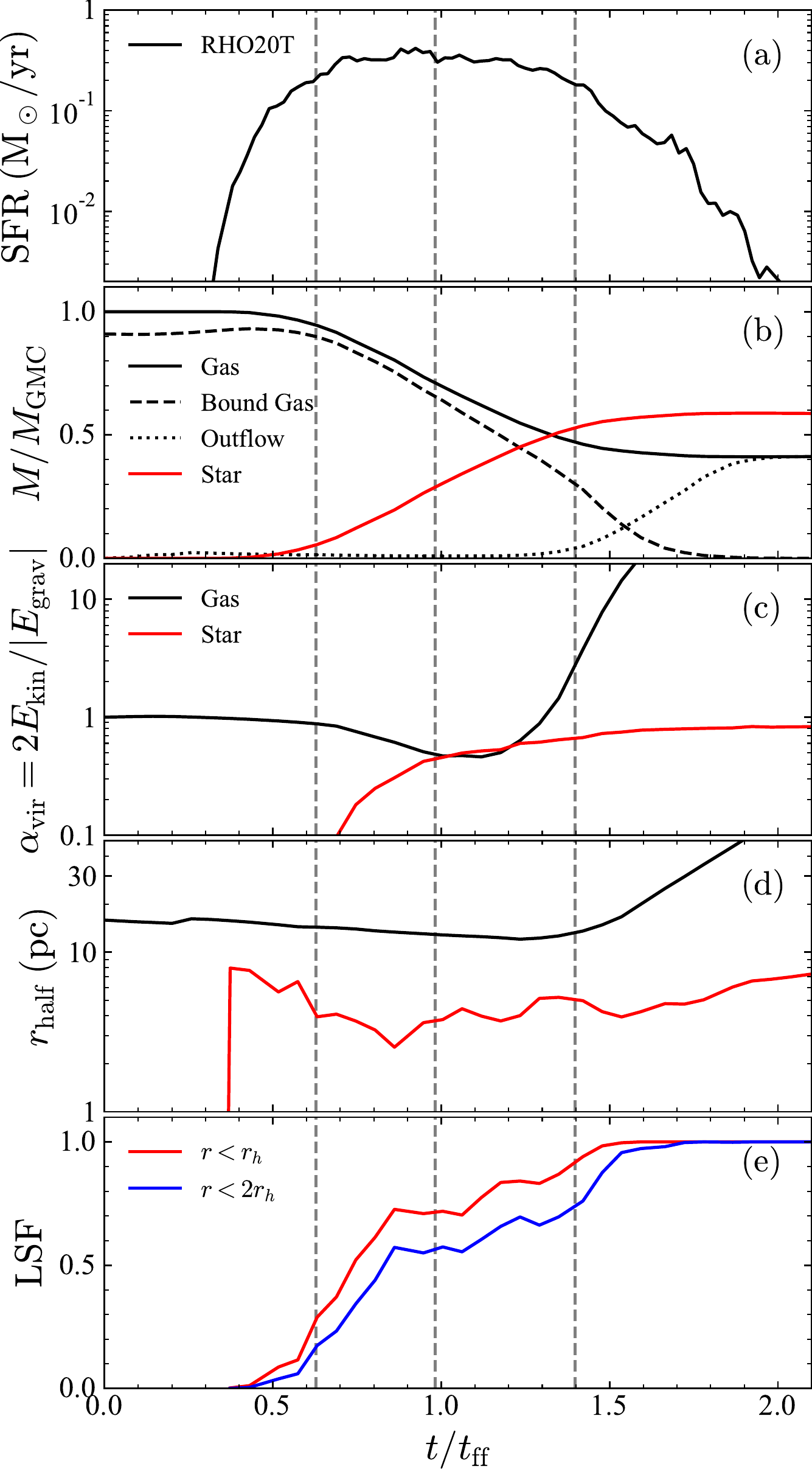}
\caption{Time evolution of various physical quantities for RHO20T run with $\fb=2$: (a) SFR, (b) gas and stellar masses, (c) virial parameters for gas and stars, (d) half-mass radii for gas and stars, and (e) LSF within one and two half-stellar mass radii. The three vertical dashed lines from left to right represent the epochs at $t_{10}$, $t_{50}$, and $t_{90}$, which are the epochs when 10\%, 50\%, and 90\% of the final stellar mass is assembled, respectively. The $x$-axis is normalized by the initial free-fall time of the cloud, $\tff=1.6$~Myr.}
\end{figure}

\subsection{Caveats of the sub-grid models}
The sub-grid model used in this paper samples star particles probabilistically with star formation criteria described in \autoref{sec:method-sf}. This is different from the more realistic sink particle approach that follows the accretion history of individual stars. We adopt this simplified star formation algorithm since the goal of this paper is to investigate the global properties of GMCs and star clusters but not to study the origin of the IMF or the detailed formation of single stellar objects.
We interpret each star particle as a single stellar population, whose feedback intensity is estimated in an IMF-averaged fashion, see \autoref{sec:method-feedback}. For the most massive GMCs in our simulations, this IMF-averaged approach captures the overall energy budget of stellar feedback \citep[see also][]{grudic_etal18}. However, it unavoidably underestimates the large variation of star formation efficiency in low-mass clouds, where a few massive stars can dominate the feedback process.
In addition, the star particles in our simulations with a fixed gravitational softening length only trace the overall mass distribution of star clusters. Therefore, the detailed dynamical evolution could in principle depend on the choice of softening length \citep[e.g.][]{bate_etal03,bate12}.
To fully capture the collisional process between star particles requires simulations that resolve the formation of individual stars over the whole mass spectrum and a more accurate gravity integrator, such as NBODY6 \citep{aarseth99}. Recent efforts have been made towards this direction \citep[e.g.][]{parker_etal15, gavagnin_etal17, wall_etal19}, however, due to the high computational cost, these simulations mainly focused on less massive GMCs and cannot explore the dynamical evolution of GMCs over a large parameter space.

Another main caveat is that we only take into account the momentum feedback from stellar winds, whose intensity is controlled by $f_{\rm boost}$. A larger $f_{\rm boost}$ is used to mimic stellar feedback from multiple feedback sources, such as stellar winds, ionizing radiation, and supernovae. However, in reality, different feedback mechanisms operate on different time-scales, at different locations, and through different physical processes. For example, ionizing radiation from massive stars strongly alters the ionizing state of the gas around massive stars and deposits both internal and kinetic energy to the surrounding medium. Moreover, for some of our simulations with very long $t_{\rm ff}$, for example SIGMA40, at the late stage of the cloud evolution when the age of the massive stars is longer than their main-sequence lifetime, core-collapse supernovae can deposit an enormous energy to the cloud and violently disrupt it. The combined effects of various feedback mechanisms will be investigated in an upcoming paper.

\section{Integrated star formation efficiency (SFE)}\label{sec:sfe}
In total, we have performed 80 GMC simulations with different masses, radii, velocity configurations, and feedback boosting factors. For all runs, we stop the hydrodynamical simulations when 99\% of the gas mass is expelled from the initial spherical regions by momentum feedback. Although different GMCs show quantitative different star formation histories and final efficiencies, the general evolutionary stages of the clouds are very similar. Here we use the RHO20T run with $\fb=2$ as an example to describe the general pattern of GMC evolution.

\autoref{fig:prj_x} shows the gas density projection of this run at four different epochs. The cloud evolution is initially governed by the turbulent velocity field which creates complex filamentary structures (upper left). After $\sim0.3\tff$, a roughly log-normal density distribution is established due to the supersonic turbulence. As turbulent energy dissipates by supersonic compression, the cloud starts to experience global contraction under self-gravity. Many subclusters are formed at the intersection of the filaments where dense gas clumps experience local runaway collapse (upper right). These subclusters move along the filaments, merge with each other frequently, and eventually form more massive subclusters. Due to momentum feedback, some gas mass is channelled outwards through low-density regions. In contrast, the high-density regions are compressed further and form young stars subsequently (lower left). When the central star cluster is massive enough so that its momentum feedback is able to counteract gravitational contraction, the majority of the gas mass is expelled from the cloud centre, causing the formation of giant wind-blowing bubbles (lower right).

\subsection{Time evolution of cloud properties}\label{sec:sfe-evolution}
In \autoref{fig:time-evolution}, we quantify the time evolution of various physical quantities of the cloud for the same run shown in \autoref{fig:prj_x}.
Panel (a) shows the star formation history of the cloud until it is fully disrupted. After the first group of stars forms at $\sim0.3\tff$, the SFR rises dramatically and peaks at around $t_{50}\sim\tff$. As momentum feedback from stellar particles clears some gas mass from the cloud centre, the SFR drops gradually. Although the whole star formation activity spans over $\sim2\tff$, the majority of the stellar mass is formed around the epoch of the star formation peak. The central 80\% of the stellar mass is assembled within 0.6-1.4~$\tff$.

The cumulative version of panel (a), the stellar mass growth history, is shown in panel (b), together with the evolution of gas mass. We split all gas mass into bound and outflow components. The bound gas is defined as the total mass of gas cells with negative (kinetic+potential) energy, while the outflow is defined as the unbound gas cells that are outside twice the initial GMC radius. It is clear that the decrease of bound gas mass is partly due to gas consumption by star formation and partly due to the increase of feedback-driven outflows. The final stellar mass reaches $\sim60$\% of the initial cloud mass, which is defined as the integrated SFE, $\epsint$.

Panel (c) shows the evolution of virial parameters for both gas ($\alpha_{\rm gas}$) and stars ($\alpha_{\rm star}$). By construction, gas is initially in virial equilibrium with $\alpha_0=1$. As the turbulent energy dissipates, the kinetic energy of the cloud decreases and the system collapses, which leads to a slight decrease of $\alpha_{\rm gas}$. The momentum feedback from stars adds kinetic energy to the gas cells and helps to increase $\alpha_{\rm gas}$ after $t\sim1.3\tff$. Eventually, the virial parameter comes back to unity. Yet momentum feedback cannot keep the cloud in a quasi-equilibrium state. $\alpha_{\rm gas}$ keeps rising and becomes much larger than unity very quickly until the majority of the gas mass is removed from the central region of the cloud. Interestingly, the virial parameter of stars, $\alpha_{\rm star}$, is always smaller than unity, suggesting that the model star cluster is sub-virial. As we will show later, this sub-virial dynamical state before gas expulsion has a dramatic effect on the formation of bound clusters in GMCs.

In panel (d), we present the evolution of the half-mass radius of the gas and stellar components of the GMC. The evolution of the half-mass radius of the gas tightly follows the evolution of its dynamical state as is described in panel (c). The size of the gas cloud first shrinks slightly due to the dissipation of the initial turbulent energy until stellar momentum feedback puffs it up. The evolution of the half-mass radius of the star cluster is more complicated. At first, stars are formed in dense gas clumps distributed over a large volume of the cloud, which leads to a relatively large initial stellar radius. As the cloud contracts, star formation activities concentrate more towards the central region and the stellar half-mass radius decreases until $t_{50}$. Later, as gas removal shallows the overall gravitational potential, the star cluster expands dynamically to reach a new equilibrium state.

Many theoretical works on gas expulsion and the formation of bound fraction suggest that, rather than $\epsint$, the local stellar fraction (LSF) is considered as an effective SFE to better probe the bound fraction of the cluster after gas expulsion \citep{goodwin09,smith_etal11,smith_etal13,farias_etal18}. The LSF is defined as the mass fraction of stellar mass within the stellar half-mass radius right before the gas expulsion:
\begin{equation}
{\rm LSF}=\frac{M_*(<r_{\rm h})}{M_*(<r_{\rm h})+M_{\rm gas}(<r_{\rm h})}.
\end{equation}
In panel (e), we show the evolution of the LSF in the simulations. Since neither the formation of stars nor the dispersion of gas happens instantaneously, the LSF changes dramatically during the course of GMC evolution. By definition, the LSF is initially zero. As star formation continues, the gas is gradually consumed and expelled from the central region, causing the increase of stellar mass and decrease of gas mass. Therefore, the LSF increases monotonically as a function of time until it reaches unity.

\begin{figure}
\includegraphics[width=\columnwidth]{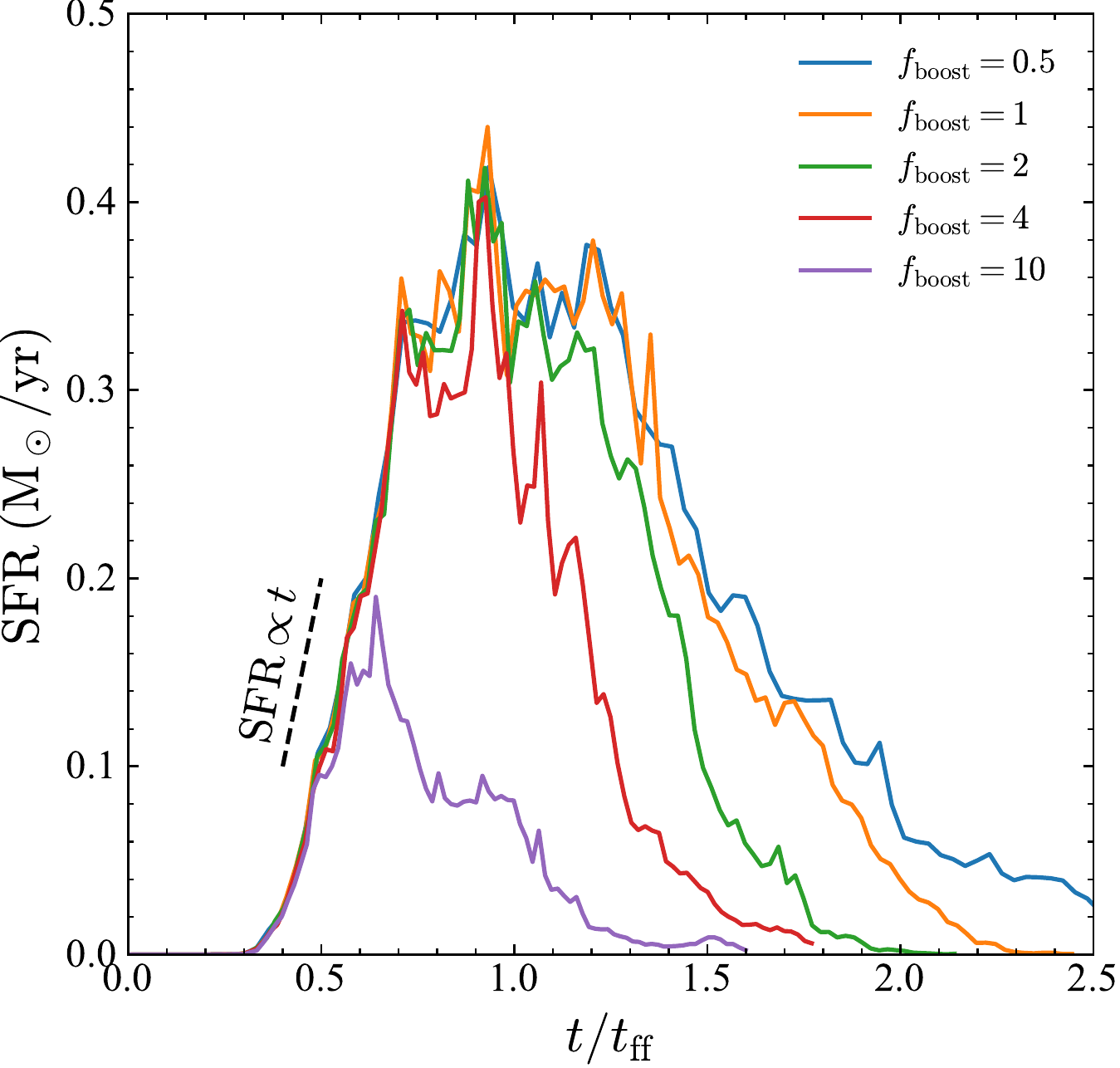}
\vspace{-5mm}
\caption{Star formation histories of RHO20T runs with five different $\fb$: 0.5, 1, 2, 4, 10. The SFR is not affected by the choice of $\fb$ during the early phase of GMC evolution. During this phase, the SFR increases linearly with time, which is shown as the dashed line: ${\rm SFR}\propto t$. Clouds start to be disrupted by stellar feedback after the SFR reaches its peak. The epoch of the peaks of star formation as well as the final SFE decreases with increasing $\fb$.}
  \label{fig:sfh_boost}
\end{figure}

\subsection{Effects of $\fb$ on the star formation history}\label{sec:sfe-fboost}
\autoref{fig:sfh_boost} shows the star formation histories of the RHO20T run with different $\fb$.
During the early stages, the total mass of young stars is so small that momentum feedback is not enough to affect the dynamical state of the cloud. Therefore, the effects of $\fb$ on the SFR is not visible and all lines overlap with each other until the SFR reaches its peak. During this period of time, the SFR presents a linear increase with time, ${\rm SFR}\propto t$. This linear time dependence is consistent with previous theoretical and numerical prediction of turbulent self-gravitating cloud with
virial parameters close to unity \citep{lee_etal15, murray_chang15, murray_etal17}.

When a sufficient fraction of gas mass is converted to stars, the momentum feedback is able to alter the overall dynamical state of the cloud (see \autoref{sec:sfe-evolution}), and eventually reverts the increasing trend of the SFR. The exact epoch of the turning point and, in turn, the final $\epsint$, is determined by the balance between the feedback intensity and gravitational contraction, which will be discussed in the next section.

\subsection{Surface density-dependent SFE}\label{sec:sfe-epsint}

\begin{figure}
\includegraphics[width=\columnwidth]{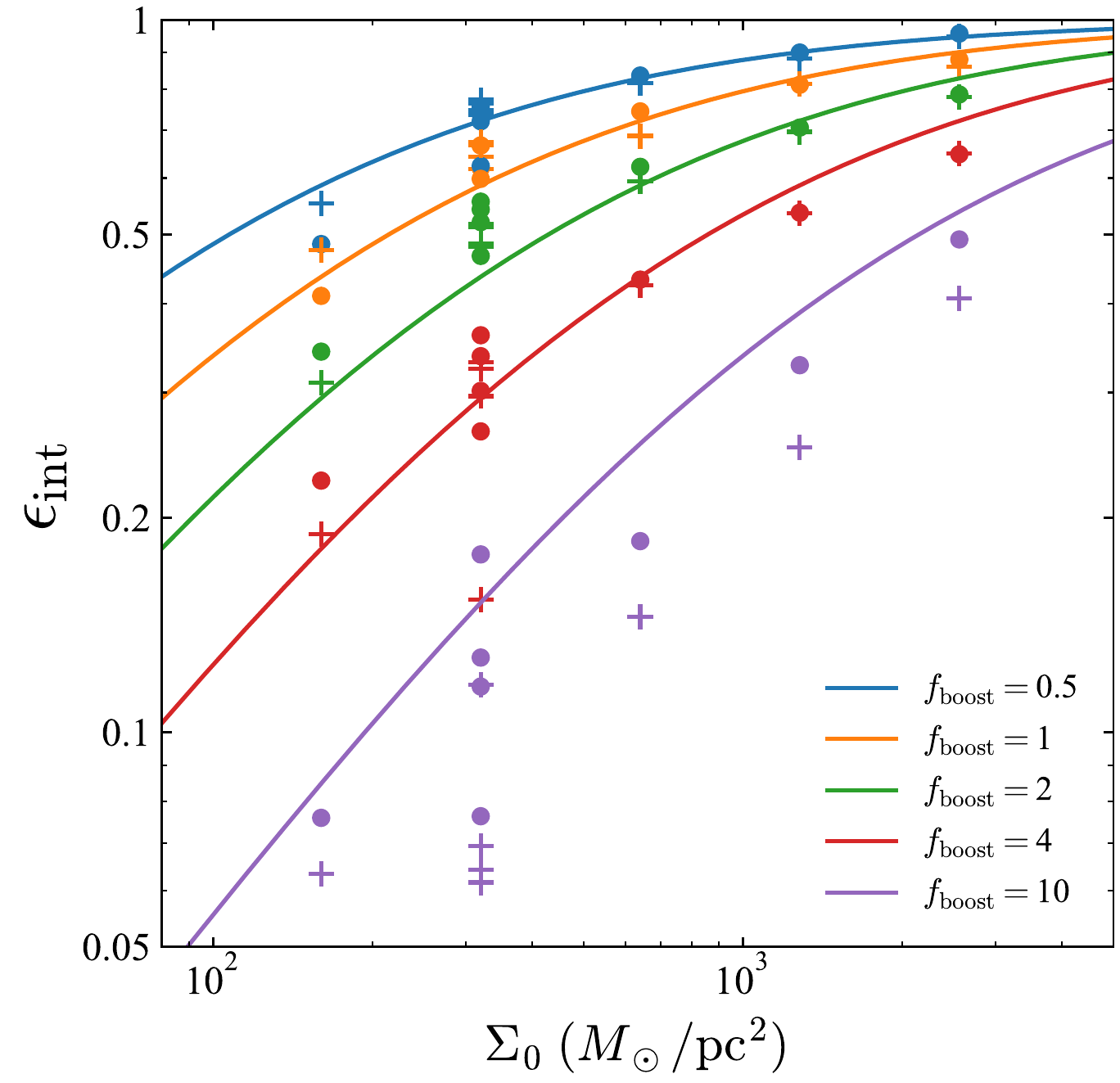}
\caption{Integrated SFE, $\epsint$, of all 80 runs as a function of initial gas surface densities, $\Sigma_0$. The solid points represent all 40 rotation-supported ``R'' runs while the crosses represent all 40 turbulent-supported ``T'' runs. Different colours represent runs using different $\fb$ (see legend for details). In general, we find that $\epsint$ increases with increasing $\Sigma_0$ and decreasing $\fb$. This trend is explained by a physical model that considers force balancing between gravitational collapse and momentum feedback. The solid lines represent the result of the physical model with best-fit parameters $\beta=1.83$ and $\dot{p}_{\rm w}=3.32\times10^{-9}{\rm cm/s^2}$ using \autoref{eq:epsint}.}
  \label{fig:epsint}
\end{figure}

For all 80 runs, we obtain the total stellar mass at the end of the hydrodynamical simulations and calculate the integrated SFE, $\epsint$. 
\autoref{fig:epsint} shows $\epsint$ as a function of the cloud initial surface density, $\Sigma_0$, for all 80 runs. We find a positive correlation between $\epsint$ and $\Sigma_0$, for a given value of $\fb$. In contrast, we do not find clear correlations of $\epsint$ with either the initial mass, radius, or volume density of the clouds. For the same GMC, runs with larger $\fb$ produce less stars and smaller $\epsint$, consistent with the results described in \autoref{sec:sfe-fboost}. Moreover, we find that rotation-supported (``R'') runs in general show a slightly higher $\epsint$ than the corresponding turbulence-supported (``T'') runs, especially when a large $\fb$ is employed. This can be explained as follows. Because of the initial rotational velocity, GMCs in the ``R'' runs first collapse to a disk-like structure whose scale height is typically smaller than the cloud radius. The formation of the thin disk allows momentum feedback to escape easier than that of the ``T'' runs, where the spherical shape of cloud is roughly maintained. This geometric effect leads to a difference of $\epsint$ by about 10-20\%.

We next build an analytical model to explain the correlation between $\epsint$ and $\Sigma_0$ by considering the force balance between gravitational contraction and gas expulsion by momentum feedback.
We assume, when the balance is achieved, the residual gas forms a thin spherical shell with a radius $r_s$. For gravitational forces, we consider the contribution from both the central star cluster of mass $M_*$ and self-gravity of the gas shell of mass $M_{\rm sh}=\Mgmc-M_*$. The gravitational force per unit area of the shell from the cluster is evaluated as
\begin{equation}
F_{\rm sh,*}= \frac{GM_*M_{\rm sh}}{r_s^2 A_s}= \frac{GM_*\Sigma_{\rm sh}}{r_s^2},
\end{equation}
while self-gravity of the gas shell is
\begin{equation}
F_{\rm sh} = \frac{\beta GM_{\rm sh}^2}{r_s^2A_s} = \frac{\beta GM_{\rm sh}\Sigma_{\rm sh}}{r_s^2},
\end{equation}
where $A_s=4\pi r_s^2$ is the surface area of the shell, $\Sigma_{\rm sh}=M_{\rm sh}/4\pi r_s^2$, and $\beta$ is the geometric factor that takes into account the anisotropic distribution of the gas shell. Note that $\Sigma_{\rm sh}$ is the surface density of the spherical shell seen from the central cluster, different from $\Sigma_0$, the cloud column density, by a factor of 4: $\Sigma_0=4\Sigma_{\rm sh}$. For a uniform density gas distribution, $\beta=0.5$. The expel force per unit area exerted onto the gas shell by momentum feedback is
\begin{equation}\label{eq:force-mom}
F_p = \frac{M_*\dot{p}}{4\pi r_s^2},
\end{equation}
where $\dot{p}$ is the momentum deposition rate per unit stellar mass. As described in \autoref{sec:method-feedback}, we use an IMF-averaged wind injection as the default setup for feedback with a boosting factor, $\fb$. In \autoref{eq:wind-lum}, the deposition rate evolves as the stellar population ages. For simplicity, in this analytical model we assume a constant momentum deposition rate per unit mass $\dot{p}=\fb \dot{p}_{\rm w}$ where $\dot{p}_{\rm w}$ is the IMF- and time-averaged value.

Force balancing between gravitational collapse and the momentum-driven wind, $F_{\rm sh, *}+F_{\rm sh}=F_p$, gives
\begin{equation}
\frac{M_*\dot{p}}{4\pi r_s^2} = \frac{GM_*\Sigma_{\rm sh}}{r_s^2} + \frac{\beta GM_{\rm sh}\Sigma_{\rm sh}}{r_s^2}.
\end{equation}
By defining variables $\Sigma_{\rm crit}\equiv\dot{p}/\pi G$ and $\Gamma\equiv\Sigma_{\rm sh}/\Sigma_{\rm crit}=\pi G \Sigma_0/4\fb\dot{p}_{\rm w}$ and assuming $r_s=\Rgmc$, the above equation can be simplified to 
\begin{equation}
(1-\epsint)\left(\frac{\beta}{\epsint}-\beta+1\right)=\frac{1}{\Gamma}.
\end{equation}
Finally, $\epsint$ can be solved as
\begin{equation}\label{eq:epsint}
\epsint = \frac{\sqrt{\Gamma^2+(4\beta-2)\Gamma+1}-(2\beta-1)\Gamma-1}{2(1-\beta)\Gamma}.
\end{equation}
For a uniform density gas shell, \autoref{eq:epsint} reduces to
\begin{equation}
\epsint = \frac{\sqrt{\Gamma^2+1}-1}{\Gamma}.
\end{equation}

For clouds with high surface density, $\Gamma\gg1$, \autoref{eq:epsint} is reduced to $\epsint\approx1-1/\Gamma$, which suggests that almost all gas mass is converted into stars before the gravitational collapse is balanced by momentum feedback. Since the mass of the gas shell is much smaller than the mass of the central star cluster, self-gravity of the gas shell is negligible and therefore $\epsint$ is independent of $\beta$. For $\Gamma\ll1$, on the other hand, the gravitational force is dominated by the self-gravity of the shell and \autoref{eq:epsint} can be simplified to $\epsint\approx\beta\Gamma$, which shows a clear $\beta$ dependence. In this case, $\epsint$ depends linearly on the cloud surface density divided by the momentum deposition rate, $\Sigma_0/\fb\dot{p}_{\rm w}$.

We fit the value of $\epsint$ for all 80 GMC simulations using \autoref{eq:epsint}, and obtain the best-fit parameters with $1\sigma$ uncertainty: $\beta=1.83\pm0.89$ and $\dot{p}_{\rm w}=(3.32\pm0.64)\times10^{-9}{\rm cm/s^2}$. As can be seen in \autoref{fig:epsint}, the analytical model is in good agreement with the simulated $\epsint$ over a large range of $\Sigma_0$ and $\fb$. We notice that the analytical model overestimates $\epsint$ for runs with $\fb=10$. This is possibly because clouds are disrupted earlier in $\fb=10$ runs than other runs and the time-averaged $\dot{p}$ is systematically higher due to the decreasing wind kinetic luminosity used in the simulations, see \autoref{eq:wind-lum}.
 
\subsection{Star formation time-scales}\label{sec:sfe-time-scale}
As described in \autoref{sec:sfe-fboost} and \ref{sec:sfe-epsint}, stronger momentum feedback changes the epoch of the peak of star formation to earlier times and reduces the final SFE. How important is the strength of feedback to the overall star formation time-scales? Can feedback be the main energy source to support the cloud and maintain a quasi-equilibrium state?
We investigate these questions here by defining several relevant time-scales that characterize the star formation activities for the simulated GMCs. First, we define the initial free-fall time of the cloud as

\begin{equation}
\tff = \sqrt{\frac{3\pi}{32G\rho_0}} \approx 1.6 {\rm Myr} \left(\frac{M}{10^5\Msun}\right)^{-1/2} \left(\frac{\Rgmc}{10{\rm pc}}\right)^{3/2}.
\end{equation}
where $\rho_0=3\Mgmc/4\pi\Rgmc^3$ is the initial volume density of the GMC. The free-fall times of all GMCs are listed in \autoref{tab:run-parameter}. We define the star formation duration as the time-scale during which the clouds form the central 80\% of their stars: $\tdur=t_{90} - t_{10}$. Following \citet{li_etal18, grudic_etal18}, we also define an age spread of star cluster as the ratio between the final stellar mass $M_*$ and the mass-weighted SFR:
\begin{equation}
\tau_{\rm spread}\equiv\frac{M_*}{<\dot{M}>}=\frac{M_*}{\int \dot{M}_*^2dt/M_*},
\end{equation}
where $\dot{M}_*$ is the instantaneous SFR. For a Gaussian-like star formation history with standard deviation $\sigma_*$, the age spread is approximated $\tau_{\rm spread}\approx2\sqrt{\pi}\sigma_*$.

\begin{figure}\label{fig:time-scales}
\includegraphics[width=1\columnwidth]{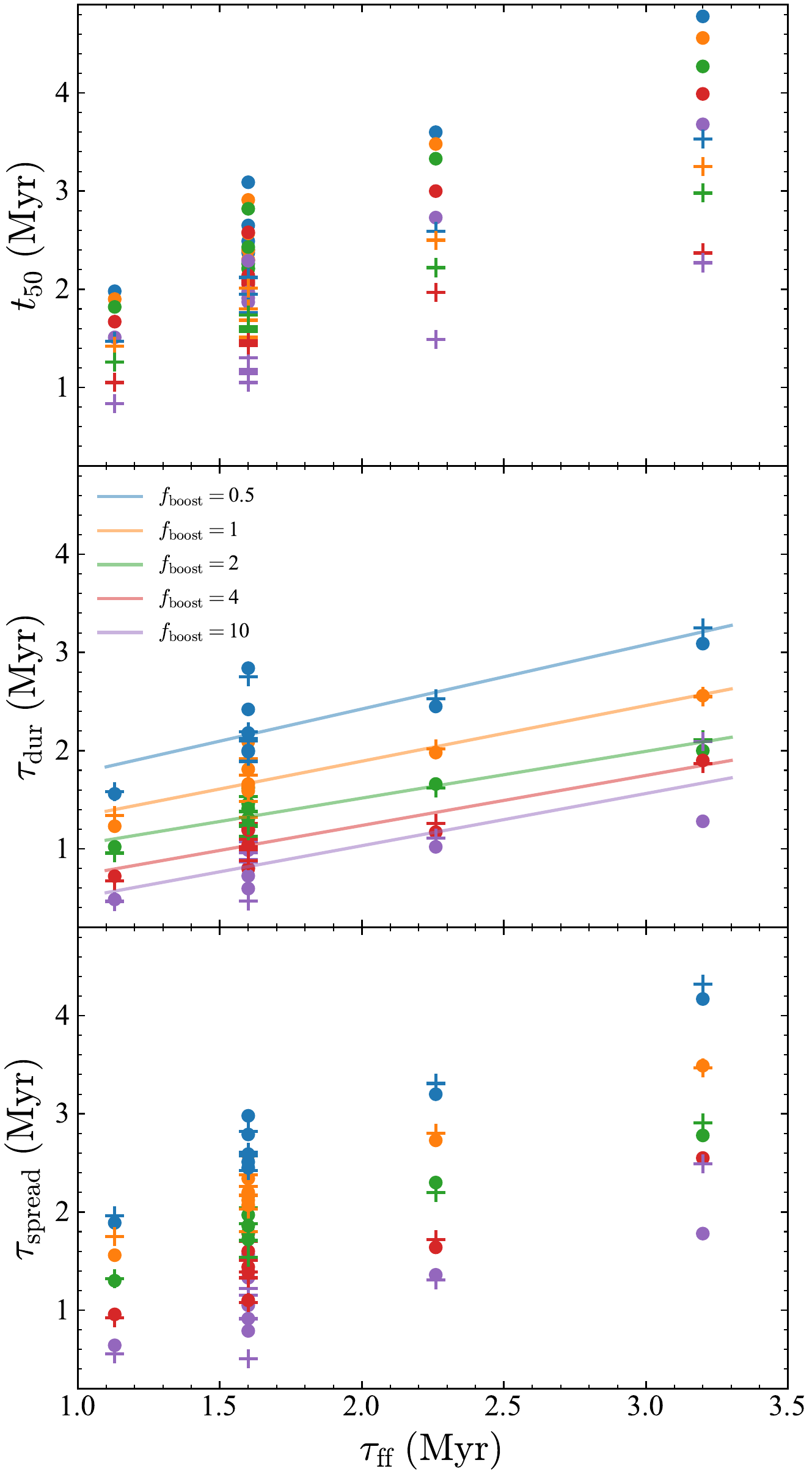}
\caption{Relevant time-scales, $t_{50}$ (upper), $\tdur$ (middle), and $\tspread$ (lower), as a function of the initial free-fall time of the GMCs for all 80 runs. $t_{50}$ roughly represents the epochs of star formation peaks, while $\tdur$ and $\tspread$ are different definitions of overall star formation durations. The same as \autoref{fig:epsint}, different colours and markers show runs with five different $\fb$ and ``R''/``T'' runs. In the middle panel, the best-fit linear relation between $\tff$ and $\tdur$ for runs with different $\fb$ is shown as solid lines. }
\end{figure}

In \autoref{fig:time-scales}, we show the central star formation epoch ($t_{50}$), star formation duration ($\tau_{\rm dur}$), and age spread ($\tau_{\rm spread}$) as a function of initial free-fall time ($\tff$) for all 80 GMCs.
In the top panel, we find a clear linear correlation between $t_{50}$ and $\tff$. We know that $t_{50}$ roughly represents the epoch of the peak of star formation because of the Gaussian-like shape of the star formation history, see \autoref{sec:sfe-evolution}. In fact, $t_{50}$ is close to $\tff$ for clouds that are turbulence-supported (``T'' runs). For ``T'' runs, the initial turbulent energy dissipates within the turbulence crossing time, which is shorter than the free-fall time of the cloud. The peak of star formation is determined by gravitational collapse of the whole cloud and is therefore similar to the free-fall time. The ``R'' runs, on the other hand, show a systematically larger $t_{50}$ than the corresponding ``T'' runs. This is because the rotation-supported cloud first collapses along the rotational axis and form a gaseous disk. The rotating disk contains more coherent motions whose kinetic energy dissipates over a longer time-scale than turbulent motions. 
Interestingly, as shown in the middle and bottom panels, the star formation duration and age spread for ``T'' and ``R'' runs do not show clear difference, which suggests that once the runaway collapse starts, the details of the initial configuration of the gas motion do not affect the subsequent star formation process.

Similar to $t_{50}$, $\tdur$ and $\tspread$ also correlate linearly with $\tff$. We perform a linear fit to the correlation between $\tff$ and $\tdur$ and find that runs with different $\fb$ show similar scalings but with different normalizations.
Although the normalization of the relations shows an anticorrelation to $\fb$, increasing $\fb$ by a factor of 20 from 0.5 to 10 only shortens the time-scale by a factor of 3 to 4. Quantitatively, this weak dependence of the momentum feedback intensity on the star formation duration can be understood as follows. In \autoref{sec:sfe-fboost}, we find a linearly increasing SFR from $t_{10}$ to $\sim t_{50}$ regardless the choice of $\fb$. Here we define ${\rm SFR}=A(t-t_{10})$, where $A$ is an arbitrary normalization. Therefore, the approximated total stellar mass is $M_{*, \rm{linear}}=5/2\int_{t_{10}}^{t_{50}}A(t-t_{10})dt=5/4A(t_{50}-t_{10})^2\sim5/16A\tdur^2\propto \tdur^2$, assuming $\tau_{\rm dur}=2(t_{50}-t_{10})$. We also obtained a correlation between $\epsint$ and $\Sigma_0$ from \autoref{eq:epsint}, which gives the final stellar mass as $M_*=\epsint M_0$.
For clouds with $\Sigma_0<<\Sigma_{\rm crit}$, the above expression can be simplified as $M_* \approx \beta\Gamma\Mgmc = \pi G \beta\Sigma_0 \Mgmc / 8\fb\dot{p}_{\rm wind}\propto\fb^{-1}$. Equating $M_{*, \rm{linear}}$ and $M_*$ gives a scaling $\tdur\propto\fb^{-1/2}$. For high surface density clouds when $\Sigma_0>>\Sigma_{\rm crit}$, $\epsint\approx1-1/\Gamma$ and therefore $\tau_{\rm dur}$ is almost independent of $\fb$. Indeed, we find that the correlation between $\tau_{\rm dur}$ and $\fb$ in the simulations scales between $\tdur\propto\fb^{-1/2}$ and $\tdur\propto\fb^0$, with a median power-law slope around -1/4. The weak dependence of star formation duration to the strength of momentum feedback suggests that the cluster formation time-scale is mainly determined by gravitational runaway collapse. Keep in mind that the turbulent velocity fields used in our simulations are initialized at the very beginning of the simulations. No subsequent turbulence driving is applied to feed in the kinetic energy after turbulence dissipation. Understanding how turbulent motions cascade from large-scale environments to the local star-forming regions and affect the long-term star formation activities requires simulations of GMCs in realistic galactic environments, which will be investigated in a future work.

\section{Bound fraction of model clusters}\label{sec:fbound}
As discussed in previous sections, momentum feedback from young stars disrupts star-forming regions, reduces the SFE, and shortens star formation time-scales. The gas expulsion and cloud disruption flatten the gravitational potential and inevitably leave some imprint on the dynamical state of the star clusters formed at the centre of the clouds.
Previous works have studied extensively the effects of gas removal on the dynamical evolution of star clusters. A simple virial analysis shows that if more than 50\% of the mass is instantaneously removed from a virialized system, the remaining mass will become gravitationally unbound \citep{hills80,mathieu83}. This conclusion is based on several assumptions: (1) the system is initially in virial equilibrium; (2) the removed mass is initially well mixed with the residual mass; (3) the mass-loss time-scale is much shorter than the dynamical time-scale of the system.
In realistic star-forming environments, all of the above assumptions are not strictly applicable. The star-forming regions are not necessarily in virial equilibrium \citep[e.g.][]{offner_etal09}. Stars are not randomly distributed within the GMCs, but are formed hierarchically within the densest molecular cores at the intersections of the gas filaments \citep[e.g.][]{smith_etal11, smith_etal13,farias_etal15,lee_goodwin16,farias_etal18}. The non-star-forming gas is expelled outward gradually rather than instantaneously \citep{geyer_burkert01,smith_etal13} and is preferentially channelled through low-density holes and tunnels rather than being removed homogeneously. Here, we explore these factors in our simulations and investigate how gas expulsion affects the bound fraction of star clusters.

\begin{figure}
\includegraphics[width=\columnwidth]{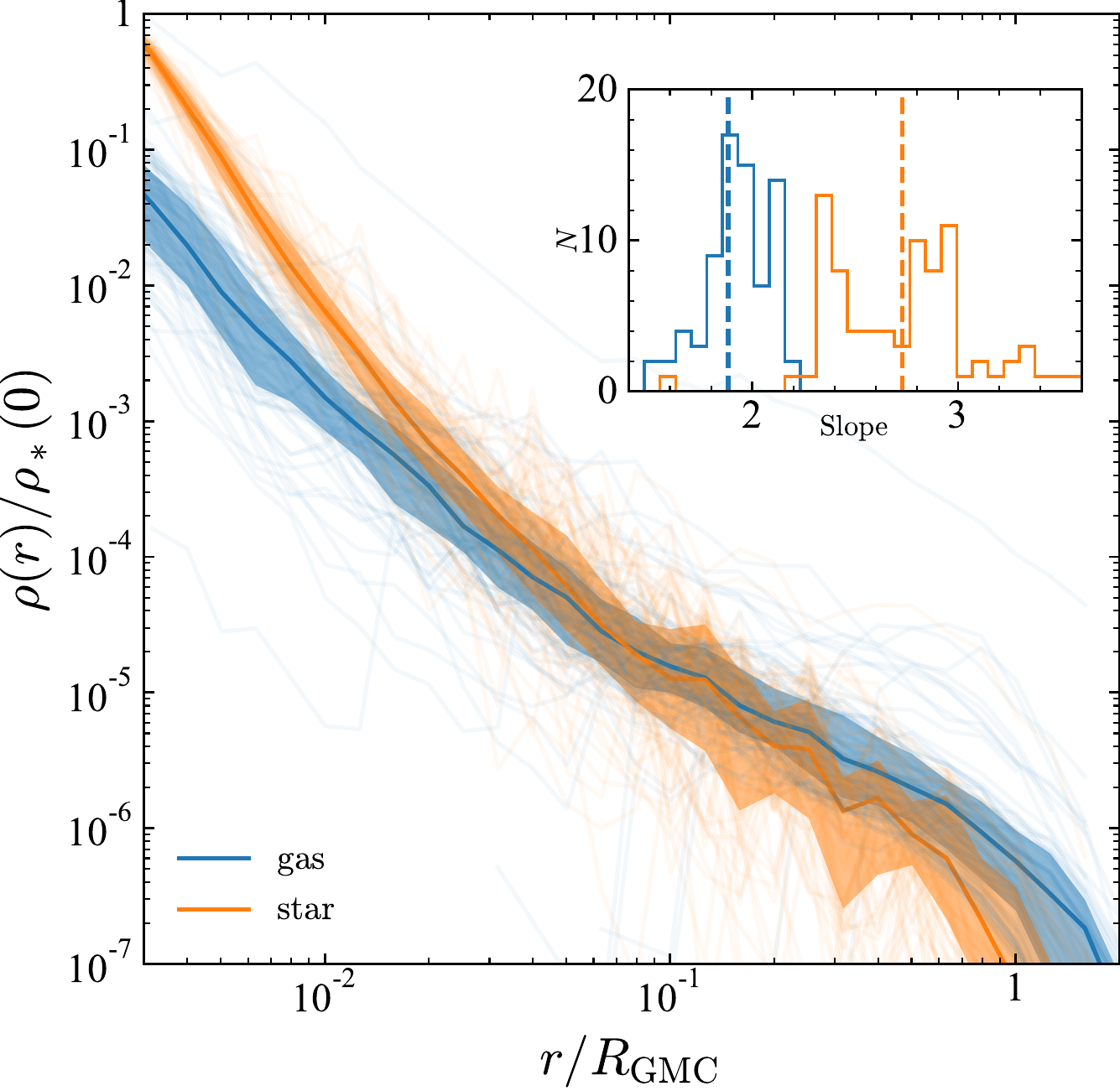}
\vspace{0mm}
\caption{Compilation of gas (blue) and stellar (yellow) density profiles for all 80 runs. The profiles are normalized to the central stellar density. Individual profiles are shown as background transparent lines, while the median and 25-75\% interquartile range are shown as solid lines and shaded regions, respectively. The inset figure shows the distribution of the best-fit power-law slopes of the gas and stellar density profiles. The vertical dashed lines are the median of distribution of the slopes for gas and stellar profiles, respectively.}
  \label{fig:profile}
\end{figure}

\subsection{Stellar and gas distribution at $t_{50}$} \label{sec:fbound-profile}

\autoref{fig:profile} shows the density profiles for both gas and stars at $t_{50}$ for all 80 runs. The profiles are centred at the location of the deepest gravitational potential, which is usually the centre of the central star clusters. Both the gas and stellar profiles are normalized to the central stellar density of the corresponding run. We find that, for most of the runs, the stellar central density is systematically higher than that of the gas, suggesting that a large fraction of gas mass in the central region of the GMCs has already been converted to stars at $t_{50}$. As a result, the gravitational potential is dominated by stars rather than gas within $\sim0.1\Rgmc$. We fit both the gas and stellar density profiles with a power-law shape, $\rho(r)\propto r^{-\gamma}$, and show the distribution of the power-law slopes $\gamma$ in the inset of the figure. The gas density profiles shows a roughly isothermal profile with a median power-law slope $\gamma\sim1.9$. It is interesting that the gas distribution for all runs converges to quasi-isothermal profiles regardless of the fact that the initial conditions are uniform density spheres.
This isothermal gas density profile is consistent with the observed radial profiles of star-forming molecular clumps \citep{mueller_etal02,palau_etal14,wyrowski_etal16,csengeri_etal17} and is thought to be a natural consequence of scale-free gravitational collapse \citep[e.g.][]{larson69, penston69, naranjo-romero_etal15, donkov_stefanov18, li18}. 
The stellar density profiles, on the other hand, are systematically steeper than that of the gas with slopes centred around 2.8 with a large variation. The steeper stellar density profiles imply more centrally concentrated star formation activities.

\subsection{Virial state of star clusters at the peak of star formation}\label{sec:fbound-virial}
In panel (c) of \autoref{fig:time-evolution}, we showed the evolution of the virial parameter of stars in the RHO20T run with $\fb=2$. We found that the model star cluster is in a sub-virial state ($\alpha_{\rm vir,*}$<1) during the course of gas expulsion from $t_{50}$ to $t_{\rm exp}$. Here we calculate $\alpha_{\rm vir,*}$ at $t_{50}$ for all 80 GMCs in order to quantify the dynamical state of the star clusters before gas expulsion. We find that $\alpha_{\rm vir,*}$ has a median value around 0.61 with a 25-75\% interquartile range 0.55-0.65, which suggests a systematic sub-virial dynamical state. This finding is qualitatively consistent with previous simulations, such as \citet{offner_etal09}, who suggest a sub-virial stellar velocity dispersion even in virialized GMCs. This is possibly because stars are preferentially formed in the densest molecular cores which on average have less velocity dispersion than the rest of the cloud. Moreover, as discussed in \autoref{sec:fbound-profile}, the stellar distribution is much more compact than that of the gas, which also helps the central cluster to remain gravitationally bound. As will be shown later, the sub-virial state has a dramatic effect on the final boundness of star clusters after gas expulsion.

\subsection{Gas expulsion time-scales vs dynamical time-scales}\label{sec:fbound-time-scale}
The dynamical response of star clusters to gas expulsion is a competition between the flattening of the gas potential that happened over gas expulsion time-scale and the energy exchange among stars that happened over the dynamical time-scale of the clusters. Previous $N$-body simulations mimic the gas expulsion process by gradually reducing the background gas potential over a given period of time \citep[e.g.][]{geyer_burkert01,baumgardt_kroupa07}. However, in realistic star-forming regions, star formation and gas expulsion happen at the same time and there is no clear separation between the two processes.

Here we define the gas expulsion time-scale as the duration from the peak of star formation to the epoch when the contribution of the gravitational potential energy from gas mass within twice the half-mass radius of the star cluster is less than 10\%, $\tau_{\rm exp}=t_{\rm exp}-t_{50}$. We also modified the potential energy threshold from 1 to 10\% and find the expulsion time-scale is not sensitive to the choice of this value. Similar to the star formation duration, we find that $t_{\rm exp}$ depends strongly on the initial free-fall time of the GMC, suggesting that gas expulsion associates well with the end of star formation. It also suggests that gas expulsion happens neither instantaneously nor much longer than the dynamical time-scale of the clouds.

\begin{figure}
\includegraphics[width=\columnwidth]{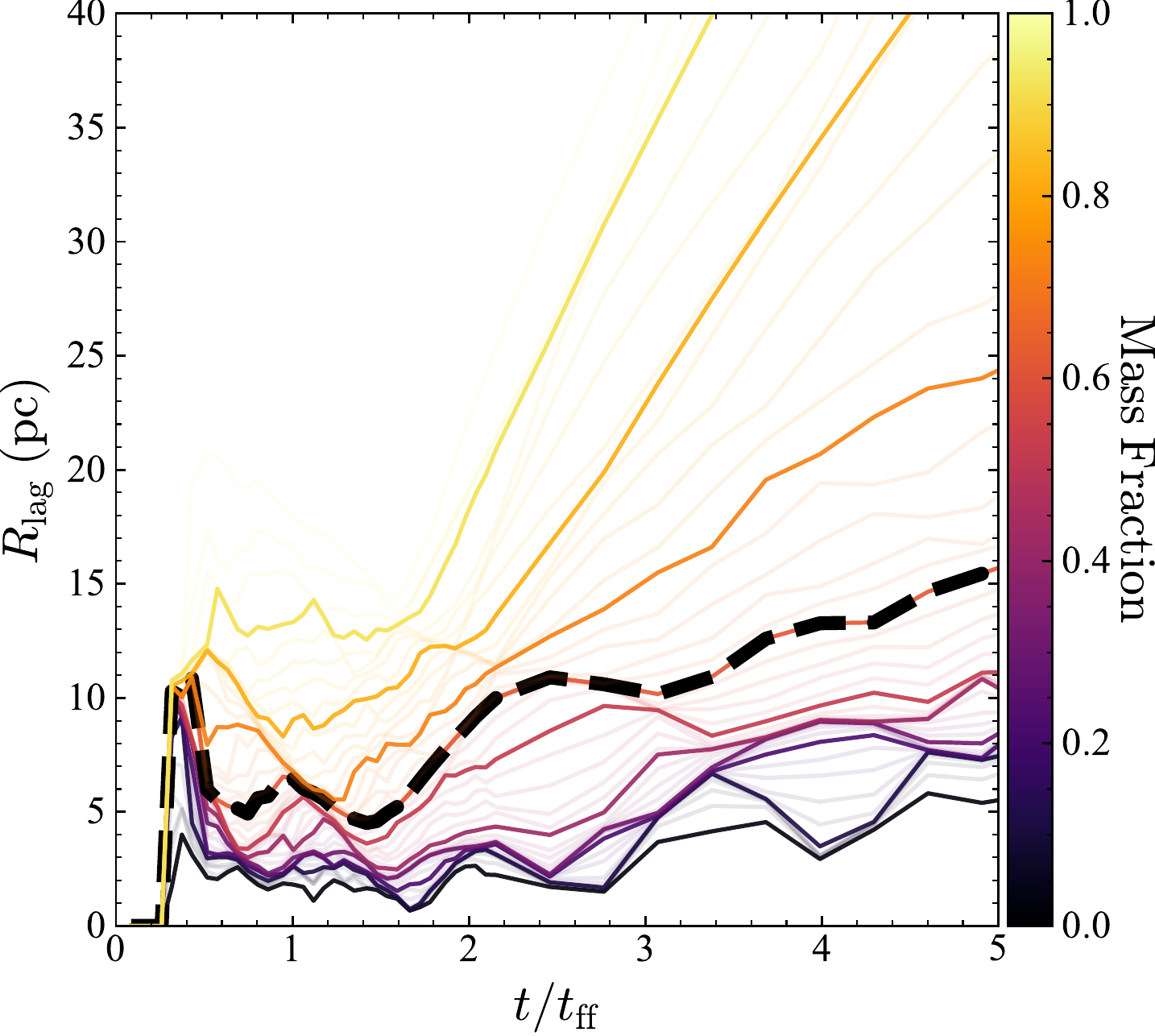}
\caption{Evolution of Lagrangian radius of star clusters formed with GMCs during gas expulsion for RHO20T run with $\fb=2$. $R_{\rm lag}$ of different mass fractions are shown as lines with different colours. Lines with less transparency are for mass fractions at every 10\%. $R_{\rm lag}$ for mass fraction that corresponds to the bound fraction estimated by the iterative method described in \ref{sec:fbound-methods} is shown as thick dashed line.}
  \label{fig:star-contour}
\end{figure}

\subsection{Calculating the bound fraction of star clusters}\label{sec:fbound-methods}
The final efficiency of star formation and all relevant time-scales investigated above are determined once the majority of the gas mass is removed from the central region of the clouds. Right after gas removal, the dynamical state of star clusters will readjust according to the changes of gravitational potential for the next couple of dynamical time-scales. 
Therefore, after momentum feedback expels more than 99\% of the gas mass out of twice the stellar half-mass radius, we stop the hydro runs, remove residual gas cells, and continue evolving the star clusters in a gravity-only mode with the same softening length of stellar particles as the corresponding hydro runs.
The simulations keep running for another two free-fall times of the GMC, $\tff$. 
We analyze the cluster bound fraction from $N$-body snapshots at different epochs and find that the bound fraction usually becomes stable after only $\sim0.5\tff$.

We adopt two methods to estimate the bound fraction from the last snapshot of the $N$-body runs.
The first and simplest way is to calculate the mass fraction of all stellar particles with negative energies. This method gives accurate results for clusters that have large bound fractions. However, for clusters with low bound fraction, simply summing up stellar particles with negative energy overestimates the bound fraction. Stars with negative energy are not guaranteed to be bound to the cluster since removing all stars with positive energy shallows the gravitational potential.
Therefore, we design a new method that removes stellar particles with positive energies and updates gravitational energies for the remaining stars iteratively. The iteration stops when all remaining stars have negative energies and the bound mass fraction of the clusters, $\fbd$, is obtained.

The second method is to use the ``Lagrangian radius'', $R_{\rm lag}$, defined as a series of radii within which the star cluster contains a sequence of fractions of stellar mass. \citet{brinkmann_etal17} suggest to use the evolution of the Lagrangian radii to determine the structural changes of the star cluster during gas expulsion. The bound fraction after gas expulsion is determined by the outermost Lagrange radius that shows a core collapse.
Figure~\ref{fig:star-contour} shows the time evolution of the Lagrangian radii for the RHO20T run with $\fb=2$. We find that the Lagrangian radii of all mass fractions decrease during the first couple of $\tff$ due to the same reason of the decrease of the stellar half-mass radius as mentioned in \autoref{sec:sfe-evolution}. After the majority of stars are formed after $\tff$, $R_{\rm lag}$ starts to increase in response to gas removal and the decrease of the total gravitational potential. The $R_{\rm lag}$ of small mass fractions shows a turnover when the central component recollapses to form the central bound clusters. 
The evolution of $R_{\rm lag}$ for mass fraction the same as the bound fraction determined by the iterative method is highlighted in the same figure. We find that the evolution of this $R_{\rm lag}$ is indeed approximately the outermost Lagrange radius that shows a turnover. We have performed the same analysis for all 80 runs and find that the iterative and ``Lagrangian radius'' methods give consistent results, confirming the eligibility of both methods. Because the outermost Lagrangian radius with turnover is determined somewhat subjectively while the iterative method always converges to an accurate result, we will only report the bound fraction that is determined by iterative methods in later sections.

\subsection{Bound fraction as a function of $\epsint$}\label{sec:fbound-epsint}

\begin{figure}
\includegraphics[width=\columnwidth]{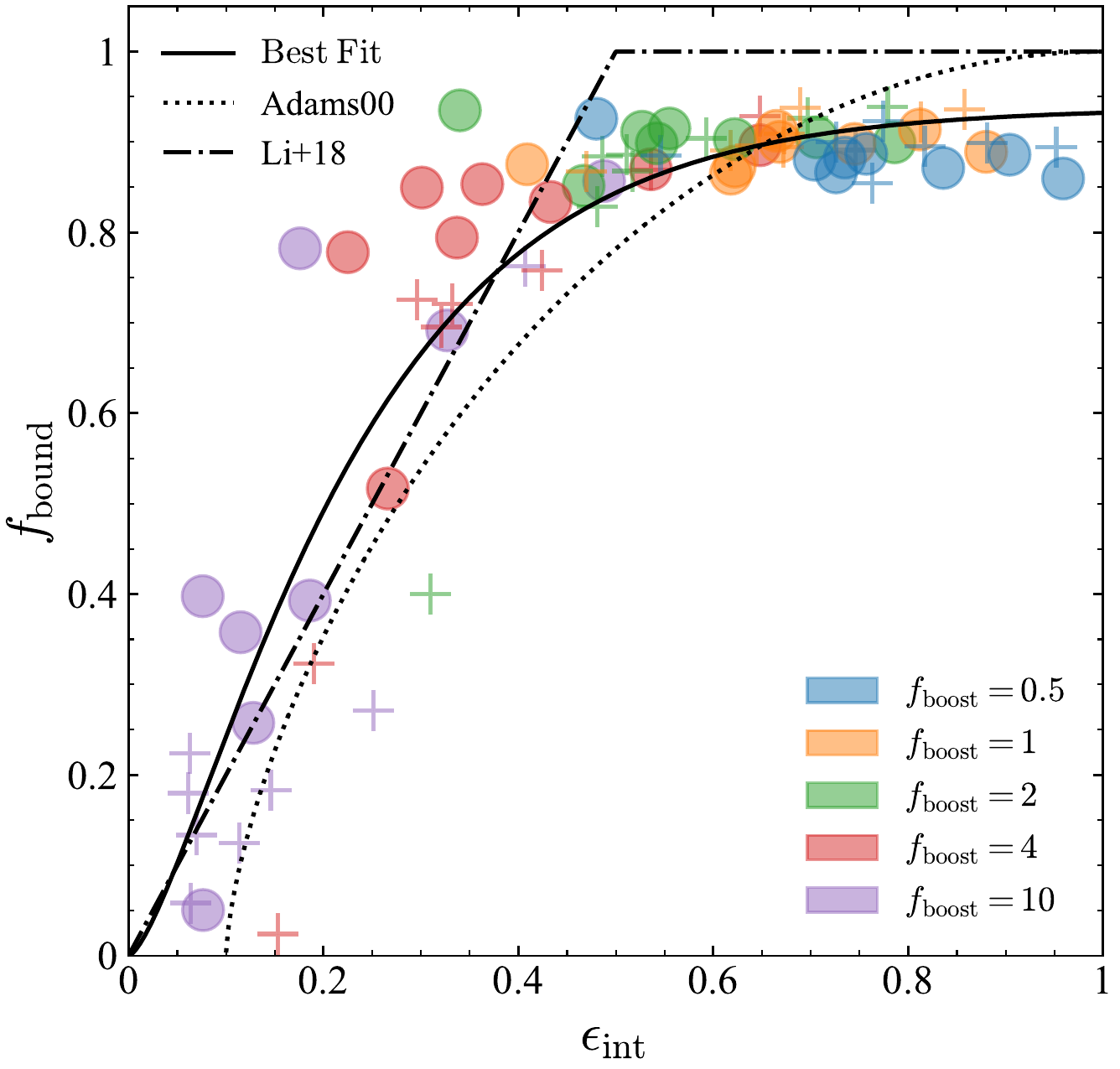}
\vspace{0mm}
\caption{Bound fraction as a function of integrated SFE for all 80 GMCs. The colour and marker styles are the same as those used in \autoref{fig:epsint}. The solid line is the best-fit model with $f_{\rm sat}=0.95$ and $\alpha_*=0.48$ using \autoref{eq:fbound-epsint}. For reference, the relationship between $\epsint$ and $\fbd$ derived from the semi-analytical model in \citet{adams00} is shown by the dotted line. The sub-grid star cluster formation model used to estimate bound cluster mass in cosmological hydrodynamical simulations in \citet{li_etal18} is shown by the dotted-dashed line.}
  \label{fig:eps-fbound}
\end{figure}

\autoref{fig:eps-fbound} shows a compilation of bound fractions for all 80 GMCs as a function of $\epsint$. The bound fractions are calculated from the last output of the $N$-body runs using the iterative method described in \autoref{sec:fbound-methods}. We find that there exists a broad range of $\fbd$ from almost completely bound to almost completely disruptive. There is an increasing trend of $\fbd$ as a function of $\epsint$. Interestingly, several runs with $\epsint<0.5$ show large $\fbd$, deviating from the simple virial analysis. For some runs with $\epsint\sim0.2$, they still form star clusters with large $\fbd>0.5$. The emergence of bound clusters in low-SFE clouds is actually in line with the findings in previous hydrodynamic simulations of GMCs \citep[e.g.][]{parker_etal15, gavagnin_etal17,farias_etal18}, although the detailed relationship between $\epsint$ and $\fbd$ shows subtle differences.

Here, we present a simple one-zone cluster model to explain the relationship between $\epsint$ and $\fbd$. Assuming stars in clusters always follow a Maxwellian velocity distribution
\begin{equation}
f(v)dv \equiv f(x)dx = \sqrt{\frac{2}{\pi}} x^2\exp{(-x^2/2)}dx,
\end{equation}
where $x\equiv\sqrt{3}v/\vrms$, $\vrms=\sqrt{3kT/m}$, and $m$ and $T$ are the average mass and ``temperature'' of the cluster. Before gas expulsion, a cluster with mass $M_*$ and radius $r_*$ has a virial parameter $\alpha_*=-2T_{*,0}/\Omega_{*,0}$, where the kinetic energy of stars $T_{*,0}=M_*\vrms^2/2$ and $\Omega_{*,0}=-G\Mgmc M_*/r_*$. Instantaneous gas expulsion flattens the gravitational potential but does not change the kinetic energy of the stars. Therefore, the potential energy of stars after gas expulsion drops according to the SFE, $\Omega_{*,1}=-GM_*M_*/r_*=\epsint\Omega_{*,0}$, while the kinetic energy does not change, $T_{*,1}=T_{*,0}=M_*\vrms^2/2$.
The escape velocity of the cluster after gas expulsion is
\begin{equation}
v_{\rm esc} = \sqrt{-\frac{2\Omega_{*,1}}{M_*}}=\sqrt{\frac{2\epsint}{\alpha_*}}\vrms.
\end{equation}

Assuming stars keep their Maxwellian distribution after gas expulsion, the bound fraction $\fbd$ can be estimated as the fraction of stars that have velocities below $v_{\rm esc}$:
\begin{eqnarray}\label{eq:fbound-epsint}
\fbd &=& f_{\rm sat}\int_0^{v_{\rm esc}}f(v)dv \nonumber \\
&=&f_{\rm sat}\int_0^{\sqrt{3}v_{\rm esc}/\vrms}f(x)dx \\
&=&\left[{\rm erf}\left(\sqrt{\frac{3\epsint}{\alpha_*}}\right)-\sqrt{\frac{12\epsint}{\pi\alpha_*}}\exp{\left(-\frac{3\epsint}{\alpha_*}\right)}\right]f_{\rm sat},\nonumber
\end{eqnarray}
where $f_{\rm sat}$ is the saturation of bound fractions. This saturation is probably due to some small fraction of substructures that are formed with low local star formation efficiencies and do not merge into the central cluster.

We fit the simulated $\fbd$ using the above relationship with two parameters, $\alpha_*$ and $f_{\rm sat}$. The best-fit values and their corresponding 1-$\sigma$ confidence intervals are $\alpha_*=0.48\pm0.02$ and $f_{\rm sat}=0.94\pm0.03$. The best-fit $\alpha_*$ is consistent with the measured virial parameters at $t_{50}$ (see \autoref{sec:fbound-virial}), suggesting that the sub-virial dynamical state of model clusters is the main driver to prevent clusters from being dissociated by gas expulsion.

\begin{figure*}
\centering
\includegraphics[width=0.49\textwidth]{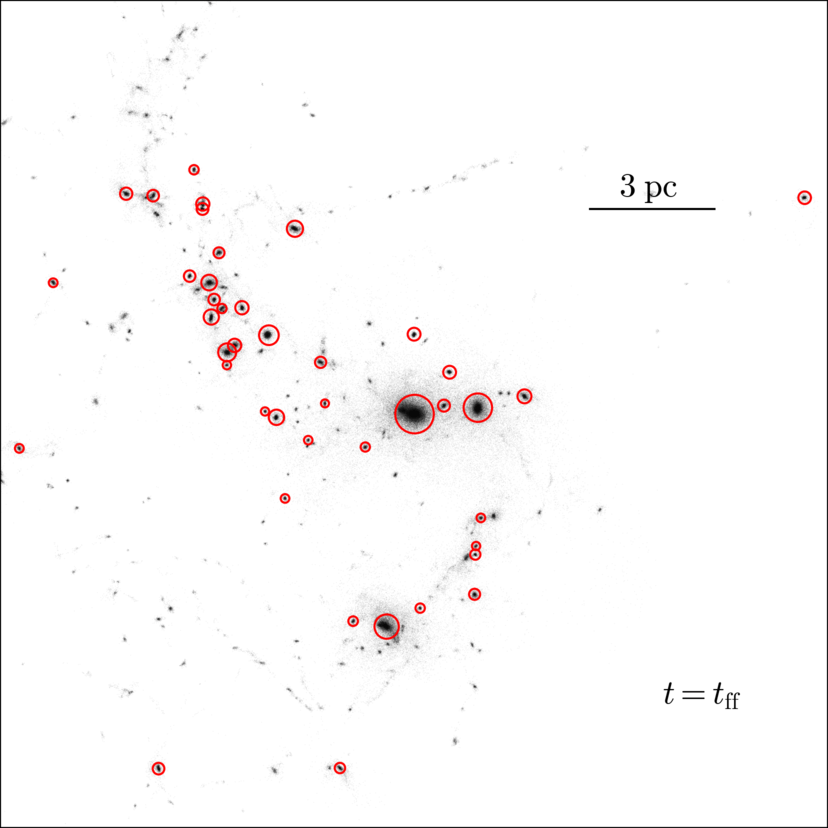}
\includegraphics[width=0.49\textwidth]{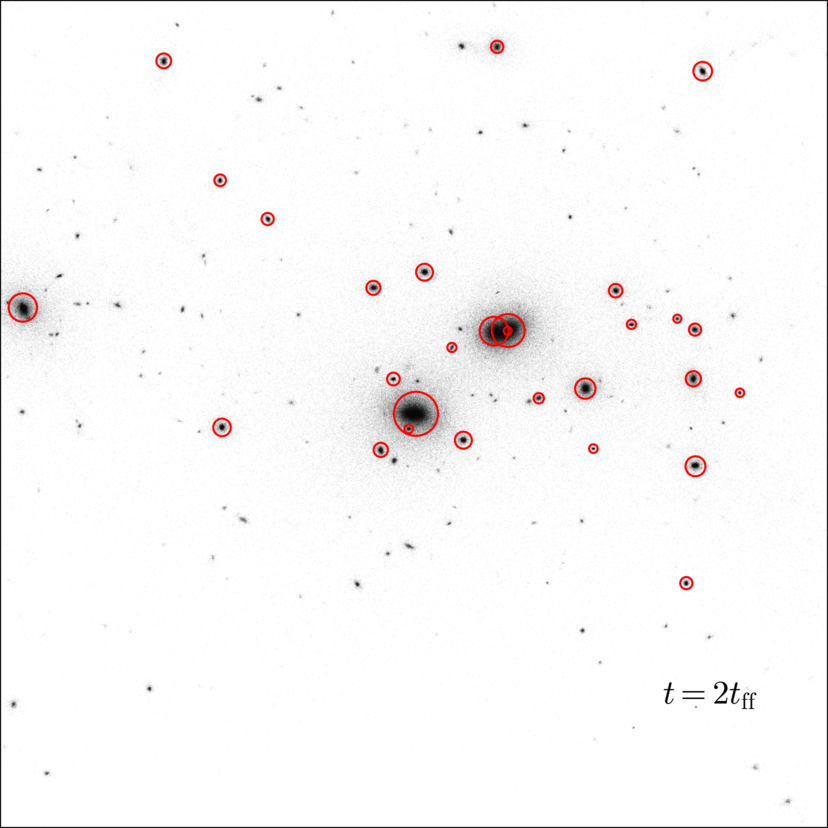}\\
\vspace{0.5mm}
\hspace{0.1mm}
\includegraphics[width=0.49\textwidth]{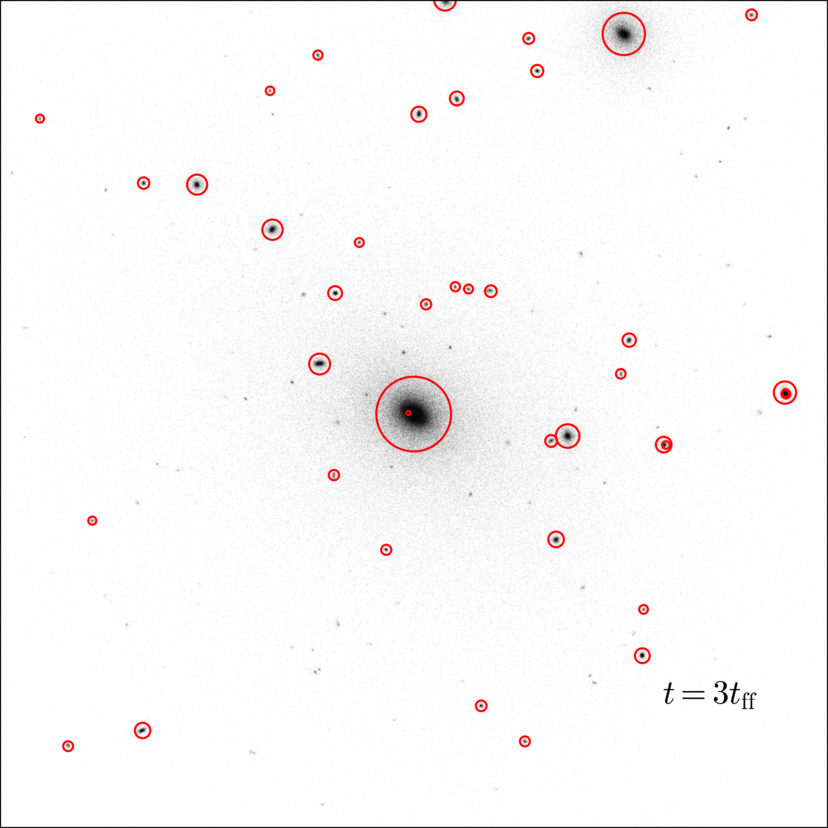}
\includegraphics[width=0.49\textwidth]{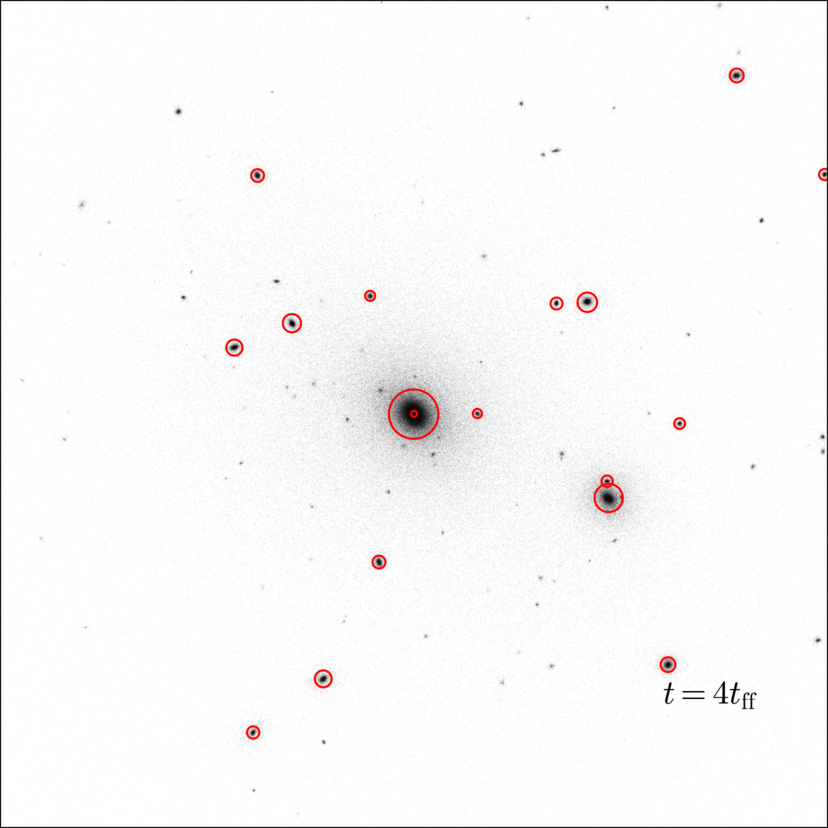}
\caption{Stellar projection plot for RHO20T run with $\fb=2$ at four different evolution epochs: 1, 2, 3, and 4 $\tff$. All plots are in the same $20\times20$ pc box centred at the centre-of-mass of the most massive subcluster. All subclusters are identified by the SUBFIND algorithm and the ones that are resolved by more than 200 stellar particles are labelled as red circles, whose radius represents the half-mass radius of the corresponding subclusters.}
  \label{fig:prj_star}
\end{figure*}

We compare our best-fit relation with the semi-analytical model developed by \citet{adams00}, who evaluates the dynamical response of star clusters to instantaneous gas expulsion based on an equilibrium cluster model with different stellar and gas density profiles. The analytical fit to his isotropic velocity distribution model gives $\fbd=(2\tilde{\epsilon}-\tilde{\epsilon}^2)^{2/3}$, where $\tilde{\epsilon}\equiv(10\epsint-1)/9$. We find that this model is roughly consistent with our simulation result but slightly underestimates $\fbd$ for $\epsint<0.5$. Moreover, we also show the $\epsint$-$\fbd$ relation used in our previous cosmological hydrodynamic simulations of a Milky Way-sized galaxy. In these simulations, a continuous cluster formation prescription is used to model the formation of individual star clusters from dense clumps resolved by parsec-scale resolutions \citep{li_etal17}. The adopted $\epsint$-$\fbd$ relation, $\fbd={\rm min}(2\epsint,1)$, determines the final bound mass of model clusters and in turn affects the cluster initial mass function, cluster formation efficiency, and the properties of evolved cluster populations \citep{li_etal18,li_gnedin18}. The piecewise function used in those cosmological simulations is in general agreement with the relationship found here.

Note that the bound fractions obtained above take into account the total bound stellar mass from not only the central star cluster but also all other surrounding substructures, which are not necessarily bound to the central cluster. In the following section, we will identify individual subclusters in the simulations, estimate the bound mass of central star clusters, and quantify its relationship to the total bound mass.

\subsection{Properties of substructures}\label{sec:fbound-sub}
Previous studies on the bound fraction of star clusters after gas expulsion usually assume an initial spherical gas and stellar distribution. However, recent observational and theoretical works suggest a hierarchical star formation scenario due to the turbulent nature of GMCs \citep[e.g.][]{elmegreen_elmegreen01,bonnell_etal03, allen_etal07, bate09, gutermuth_etal09, bressert_etal10, girichidis_etal11, maury_etal11}.
In \autoref{fig:prj_star}, we show the stellar particle distributions in the $x$-$y$ plane for the RHO20T run with $\fb=2$ at four different epochs. At $t=\tff$, the stellar distribution follows well with the gas distribution and subclusters are distributed along the filamentary structures. At $t=2\tff$ when the majority of the gas mass has already been pushed out of the central region, we find that some subclusters spiral into the centre of the GMCs due to gravitational attraction and, at the same time, merge with each other, and form more massive subclusters. Some of the most massive subclusters show a non-spherical shape because of recent mergers. At $t=$~3-4~$\tff$, dynamical evolution after gas expulsion and violent relaxation during mergers erase the memory of the turbulent configurations of the gas cloud and help circularize the central star clusters. Some small substructures with high bulk velocities escape from the central region and never return to the central cluster. 

To fully analyze the behavior of subclusters and quantify the central cluster properties, we identify bound substructures in the $N$-body simulations by adopting the SUBFIND algorithm \citep{springel_etal01}. The star particles are first linked into friends-of-friends (FOF) groups with separation less than 0.17 of the mean particle separation. For each FOF group, the SUBFIND algorithm is applied to identify all bound subclusters. We report bound substructures that are resolved by at least 200 stellar particles and assign the most massive bound subcluster as the central cluster. In most cases, the central cluster sits very close to the centre of the GMC and is surrounded by other less massive subclusters.

\begin{figure}
\includegraphics[width=\columnwidth]{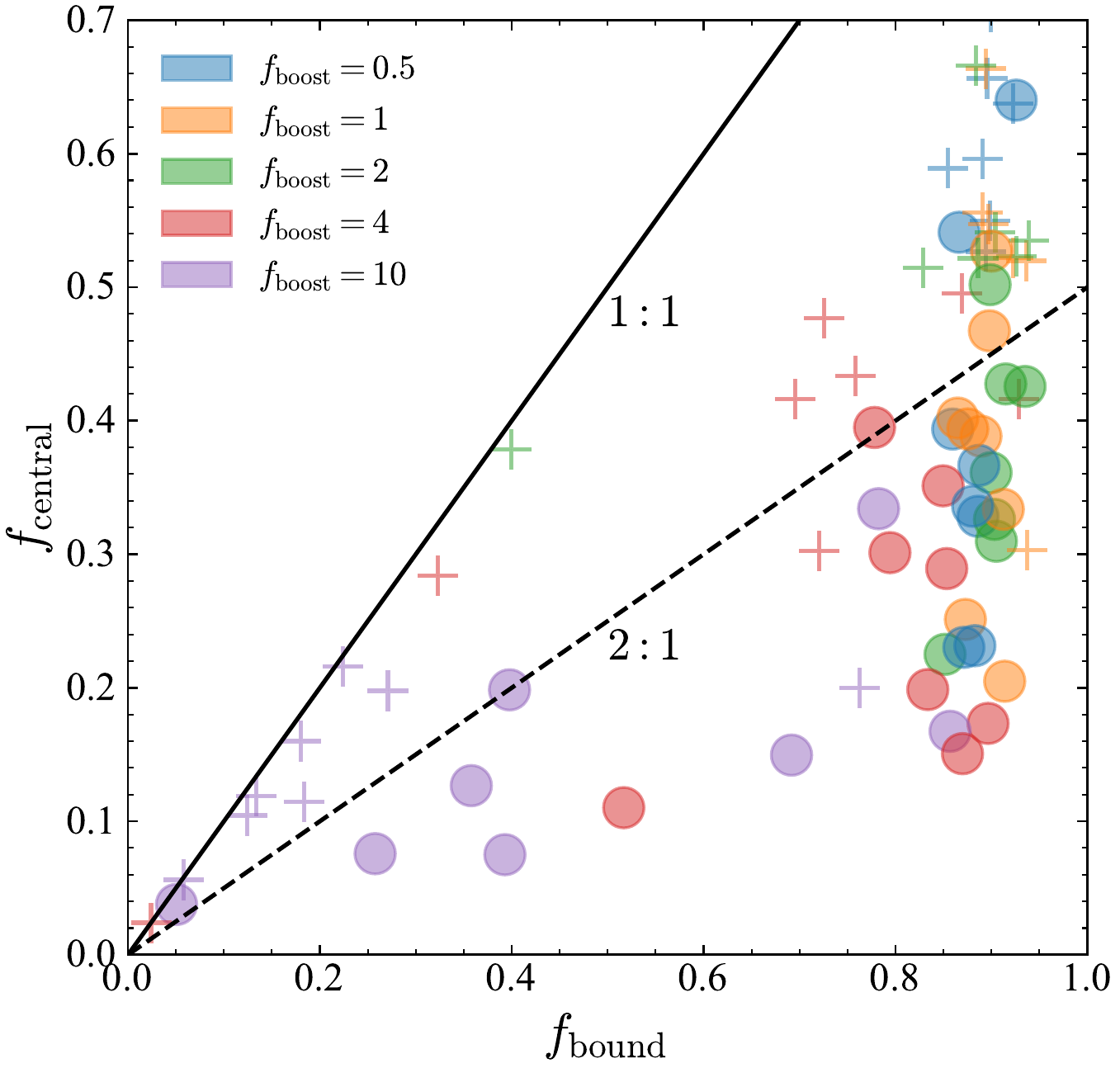}
\vspace{0mm}
\caption{Mass fraction of central cluster, $f_{\rm central}$, as a function of bound fraction, $\fbd$, for all 80 simulated GMCs. $f_{\rm central}$ is defined as the ratio of the mass of the most massive subcluster to the total stellar mass in the GMC. The colour and marker styles are the same as those used in \autoref{fig:eps-fbound}. The solid and dashed lines show the $1:1$ and $2:1$ ratio between $\fbd$ and $f_{\rm central}$.}
  \label{fig:fbound-sub}
\end{figure}

We find that the number of subclusters in the ``R'' runs is systematically larger than that in the corresponding ``T'' runs, although the total bound stellar masses are similar. In \autoref{fig:fbound-sub}, we examine the relationship between the bound fraction, $f_{\rm bound}$, and the ratio of the mass of the central cluster to the total stellar mass, $f_{\rm central}$. In most cases, the central clusters dominate the total bound mass of the system. We also find that $f_{\rm central}$ is systematically higher for ``T'' runs than for ``R'' runs. Statistically, more than 50\% of the bound stellar mass is contributed by the central clusters for ``T'' runs. This fraction is smaller for ``R'' runs but it exhibits a large scatter. As we discussed in the previous sections, GMCs in the ``R'' runs first collapse along the z-axis and form a gas disk. Therefore, many subclusters, that are formed in the dense clumps within the disk, obtain a similar bulk rotational velocity and are less likely to merge with each other than those formed in the ``T'' runs. The large number of subclusters in the ``R'' runs also implies that the stellar mass is distributed more broadly across different subclusters. Therefore, the central clusters in the ``R'' runs are less massive than those in the corresponding ``T'' runs. The exact hierarchical structure of the subclusters in the ``R'' runs depends strongly on the initial setup of the turbulent velocity fields, making the prediction of the mass of central star clusters less promising. This difficulty is reflected in the very large scatter of $f_{\rm central}$ for clouds with $\fbd>0.8$.

It should be noted that recent observations of cluster formation efficiency, defined as the fraction of stars formed in bound clusters, identify centrally-concentrated clusters of spherical shape as bound clusters \citep[e.g.][]{adamo_etal15}. This cluster selection method suggests that only the central clusters (and maybe other most massive subclusters) formed in GMCs are identified as bound star clusters when estimating cluster formation efficiencies from given galaxy patches. When interpreting the relevant observational correlations, such as the positive correlation between the SFR surface density and the cluster formation efficiency, through physics models \citep{kruijssen_etal12,li_etal18,li_gnedin18}, the effect of substructures that are not bound to the central star clusters needs to be considered and taken with caution.

\section{Summary}\label{sec:summary}
We have performed a suite of three-dimensional hydrodynamic simulations of turbulent GMCs using the moving-mesh code \textsc{Arepo} with self-gravity, explicit cooling, star formation and momentum stellar feedback. We survey a large range of GMC masses and radii, and investigate the physical origin of the large variation of intrinsic SFE, $\epsint$. After the gas clouds are fully disrupted by stellar feedback, we follow the subsequent dynamical evolution of star clusters formed within the GMCs with $N$-body simulations.
Below, we summarize our key conclusions:

\begin{itemize}
\item All simulated GMCs follow an initial linear increasing SFR before stellar momentum feedback disperses the whole cloud. The accelerating star formation activity leads to a superlinear stellar mass growth with time, $M_*\propto t^2$. This superlinearity is consistent with previous theoretical expectations of the gravitational runaway collapse of turbulent clouds.
\item Momentum feedback from stellar particles adds kinetic energy to their ambient gas cells, inflates the virial parameters and radius of the clouds, drives outflows through low-density regions, and finally creates a large cavity at the centre of the clouds. The peak epoch and final efficiencies of star formation decrease with increasing strength of momentum feedback.
\item $\epsint$ does not depend on the initial mass or radii of the clouds, but depends strongly on initial cloud surface density. This dependence is successfully explained by an analytical model that considers force balancing between gravitational collapse and momentum output from stellar particles. The model predicts $\epsint\approx1-\Sigma_{\rm crit}/\Sigma_0$ for clouds with high surface density while $\epsint\propto\Sigma_0/\fb\dot{p}_{\rm w}$ for clouds with low surface density.
\item The duration of star formation in simulated GMCs is close to the initial free-fall time of the clouds, suggesting that the cluster formation time-scale is mainly determined by gravitational runaway collapse.
The duration decreases with increasing feedback intensity, although the dependence is weak: $\tau_{\rm dur}\propto \fb^{-1/4}$. 
\item The model star clusters are assembled hierarchically. Subclusters are formed at the many density peaks across the GMCs controlled by the initial turbulence configuration. The subclusters move along the filamentary structures and merge with each other frequently. About 50\% of the mass of subclusters is merged to form the most massive central clusters, but there are always a small fraction of subclusters that are unbound to the system and fly apart from the central clusters.
\item The gas density distribution rearranges from an initial uniform density sphere to an isothermal profile with a power-law slope $\gamma\sim2$. At the peak of star formation, the stellar density profiles are systematically steeper than that of the gas with a power-law slope $\gamma\sim2.8$. The steeper stellar profiles suggest a fast conversion of gas to stars and gas expulsion by stellar feedback at the centre of GMCs.
\item The model star clusters are always in a sub-virial state with a gradually increasing virial parameter as star formation continues. Interestingly, right before gas expulsion, the virial parameters of all simulated star clusters show a consistent value around $\alpha_{\rm vir,*}\approx0.6$.
\item Due to the steep density profiles and sub-virial dynamical state of model clusters, clouds with low $\epsint$ (0.2-0.4) are still able to form clusters with relatively high bound fractions (0.3-0.8). The bound fraction of model clusters, $\fbd$, is a continuously increasing function of the integrated SFE, $\epsint$. This relation is explained by a physical model that takes into account the mass fraction of stars with velocities below the escape velocity of a sub-virial system that obeys Maxwellian velocity distribution. The best-fit virial parameter of this model is around 0.5, consistent with the values obtained directly from the simulations.
\end{itemize}

\section*{Acknowledgements}
We thank the anonymous referee for detailed comments and suggestions.
We thank Volker Springel for giving us access to \textsc{Arepo}. We are grateful to Peter Behroozi, Andreas Burkert, John Forbes, Nick Gnedin, Mike Grudic, Lee Hartmann, Jens Kauffmann, and Vadim Semenov for insightful comments and suggestions.
MV acknowledges support through an MIT RSC award, a Kavli Research Investment Fund, NASA ATP grant NNX17AG29G, and NSF grants AST-1814053 and AST-1814259. FM is supported by the Program ``Rita Levi Montalcini'' of the Italian MIUR. OG acknowledges supported through NSF
grant 1412144. The simulations of this work were run on the Harvard Odyssey clusters and the Comet HPC resource at San Diego Supercomputer Center as part of XSEDE through TG-AST170042 and TG-AST180025.






\newpage

\appendix

\section{Tests of wind-blowing bubbles}\label{sec:appendix-wind}
We test the momentum deposition algorithm used in this paper by performing idealized simulations of wind-blowing bubbles. In this test, a stellar particle is located at the centre of a uniform density box with size 20~pc and total gas mass 320~$\Msun$.
The gas mass is initially resolved by $128^3$ gas cells. The central star deposits its wind material with a constant mass-loss rate $\dot{m}_{\rm w}=10^{-5}\Msun/$yr at a fixed velocity $v_{\rm w}=500$~km/s. No self-gravity or cooling is used in this test in order to compare to analytical solutions of the evolution and internal structure of the wind-blowing bubble derived by \citet{weaver_etal77}.

\begin{figure}
\includegraphics[width=1\columnwidth]{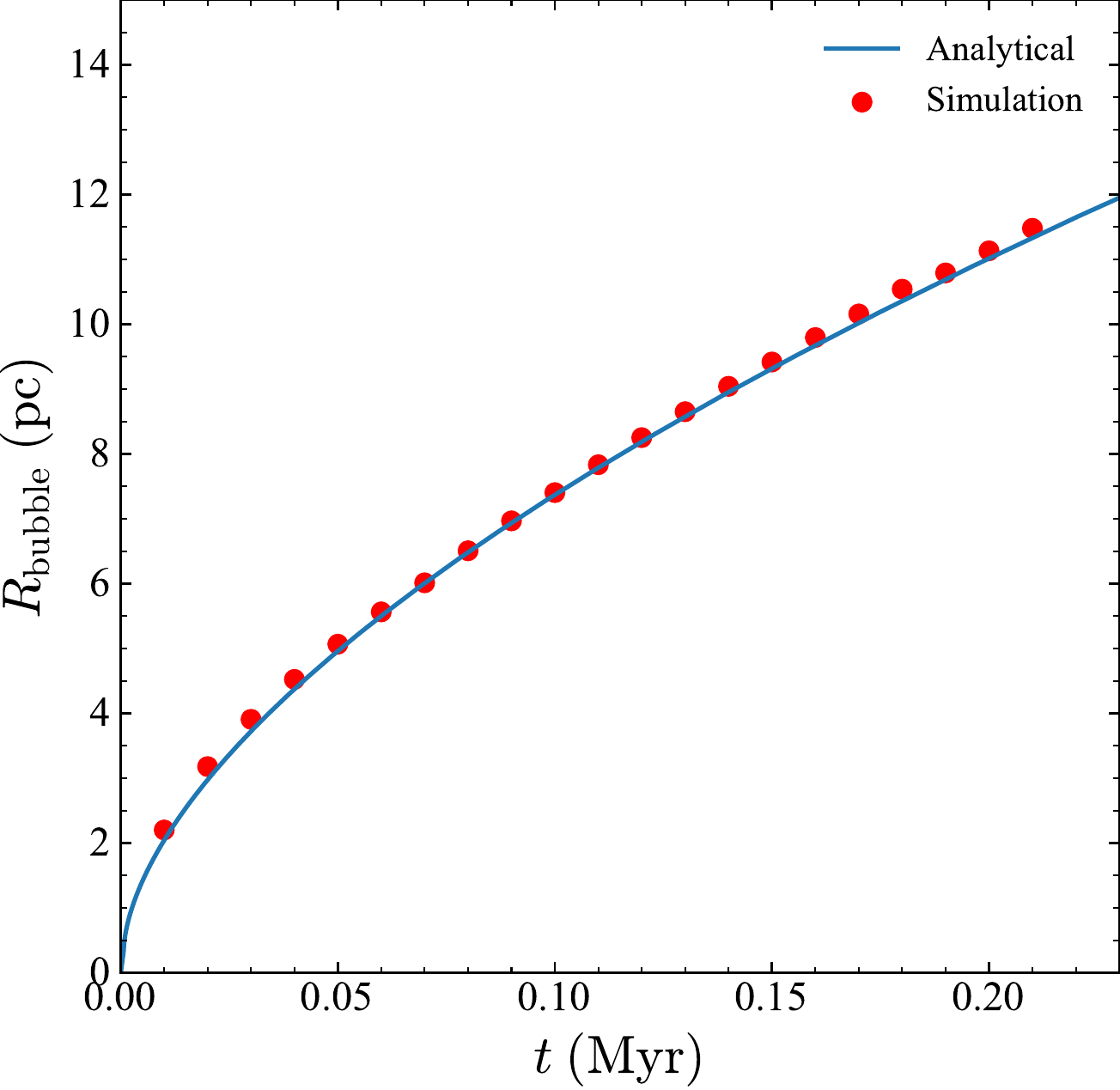}
\vspace{-2mm}
    \caption{Time evolution of the radius of wind-blowing bubble from the test simulation (red dots). The analytical solution derived from \citet{weaver_etal77} is overplotted as blue solid line for comparison.}
  \label{fig:wind-radius}
\end{figure}

\begin{figure}
\includegraphics[width=1\columnwidth]{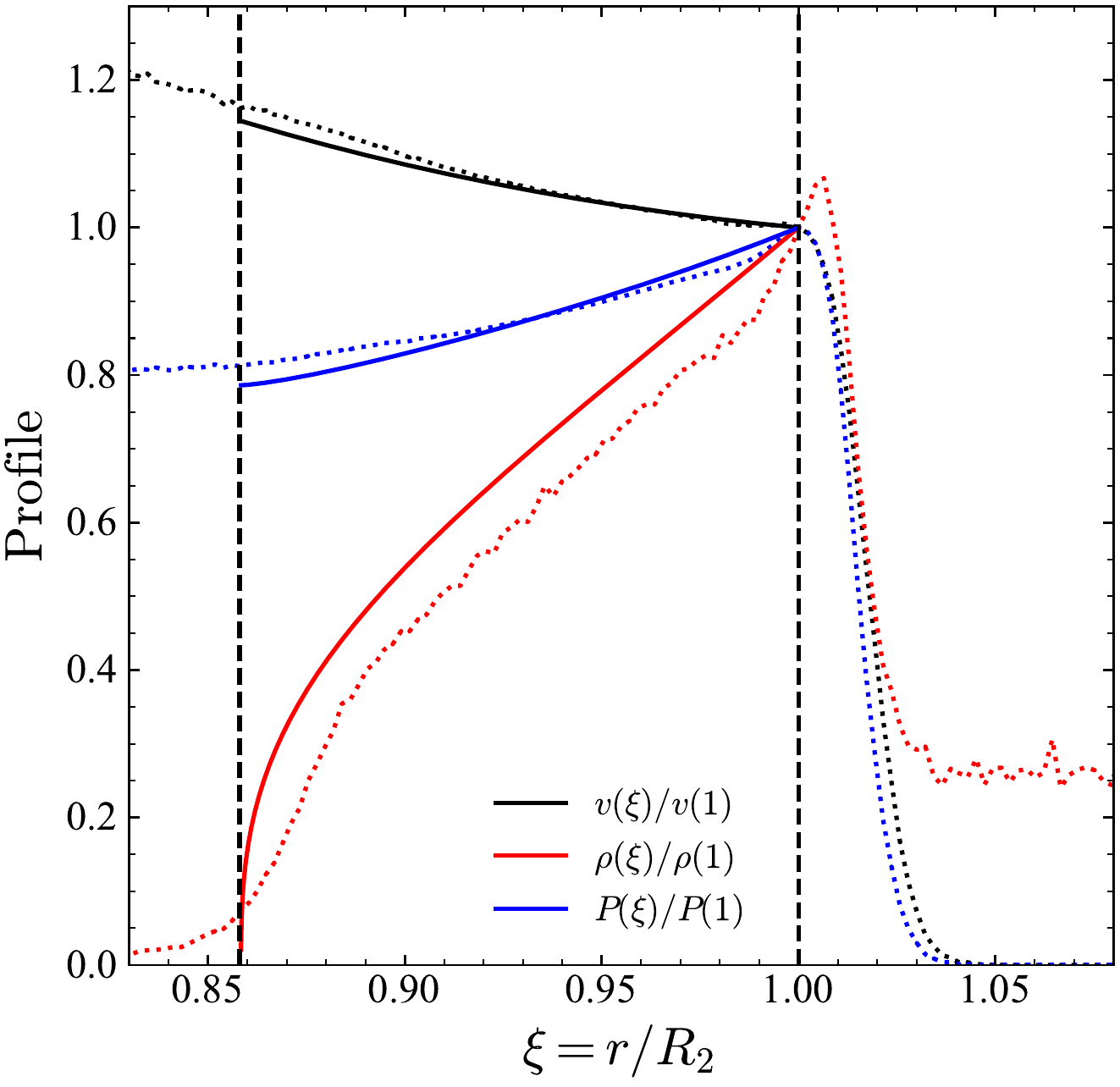}
\vspace{-2mm}
    \caption{Velocity (black), density (red) and pressure (blue) profiles within the wind-blowing bubble. All physical quantities, $v, \rho, P$, are normalized to the values at the shock front for convenience. The x-axis is normalized to the outer shock radius $R_2$. The simulation result is shown as dotted lines, while the analytical expression from \citet{weaver_etal77} is shown in solid lines for comparison. }
  \label{fig:wind-profile}
\end{figure}

The bubble first experiences free expansion with constant wind velocity until the mass of the swept-up material, $\frac{4}{3}\pi(v_{\rm w}t)^3\rho_0$, is comparable to the mass of the wind ejecta, $\dot{m}_{\rm w}t$. The initial free expansion phase only lasts for several hundred years and is followed by an adiabatic expansion phase. The time evolution of the bubble is $R_{\rm bubble}\approx0.88(\dot{m}_{\rm v}v_{\rm w}^2/2\rho_0)^{1/5}$, see Equation~(5) in \citep{weaver_etal77}.
\autoref{fig:wind-radius} shows the time evolution of the wind-blowing bubble from the numerical test. The edge of the bubble is identified as the densest gas shell surrounding the stellar particles. The radius of the bubble is calculated as the mass-weighted shell radius.
In addition to the time evolution of the gas shell, we also examine the internal structure of the bubble. \autoref{fig:wind-profile} shows the density, velocity, and pressure profiles of the shocked interstellar gas at $t=0.2$~Myr. All profiles are normalized to the values at the outer shock of radius $R_2$. The analytical solution of the self-similar adiabatic flow within the shocked medium is obtained by numerically solving Eq.~(6)-(8) in \citet{weaver_etal77}. We confirm that the momentum deposition algorithm used in our simulations reproduces the analytical time evolution and internal structure of the wind-blowing bubble with high precision.

\section{Convergence to momentum deposition methods}\label{sec:appendix-wind-gmc}
In \autoref{sec:method-feedback}, we describe various algorithms for wind deposition to the ambient medium. Here, we test how different algorithms affect the star formation activities of the GMCs.
The test runs use an identical simulation setup as the main GMC simulations described in \autoref{sec:methods} except wind feedback. We use three different weights to deposit momentum flux to neighbouring cells: volume, mass, and solid angle from star particles. We also test an alternative algorithm that injects stellar winds in the form of pure thermal energy, rather than momentum.

\begin{figure}
\includegraphics[width=1\columnwidth]{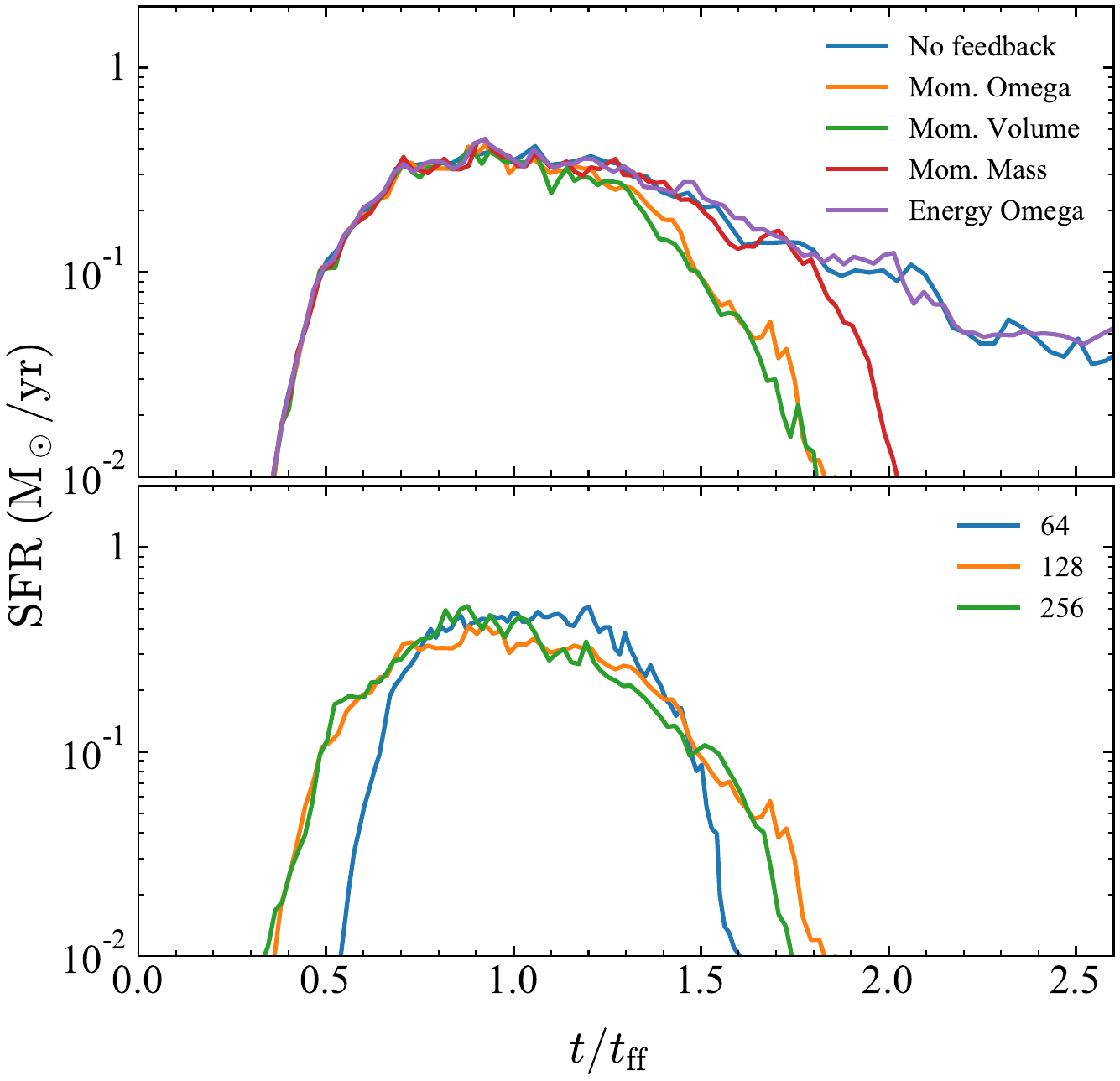}
\vspace{0mm}
\caption{\textbf{Upper panel}: SFH of GMCs with different wind feedback mechanisms: without feedback (blue); energy feedback (purple); and momentum feedback with weights of solid angle (orange), cell volume (green), and cell mass (red). \textbf{Lower panel}: The SFH of GMCs with different mass resolutions (lower). The blue, orange, and green lines show the simulations with the number of resolution elements $64^3$, $128^3$, and $256^3$, respectively.}
  \label{fig:sfh-wind}
\end{figure}

In the upper panel of \autoref{fig:sfh-wind}, we show the star formation histories of the GMCs using different wind deposition algorithms from the same initial condition RHO20T with $\fb=2$. We find that wind feedback in pure energy form is not able to disrupt the cloud and the star formation history is almost the same as that of the run without wind feedback. The failure of this energy deposition algorithm seems inconsistent with the result found in \citet{rogers_pittard13}, who used thermal energy injection to simulate wind feedback and obtained significant gas outflows from the central turbulent cloud. Note that, in their simulations, self-gravity is not included and the turbulent cloud never collapses to higher density. The maximum number density reached in their simulations is about $\sim10^4{\rm cm}^{-3}$. In contrast, our simulations track gravitational runaway collapse of dense clumps until they are converted to stars. In fact, our simulations always form much higher density clumps, $>10^{10} {\rm cm}^{-3}$. The thermal energies deposited into these dense environments suffer significant radiative cooling and therefore make the energy deposition inefficient. To capture the correct thermal dynamics of the adiabatic phase of the wind-blowing bubbles, the cooling radius needs to be resolved: $R_{\rm cool}\propto n^{-3/7}l_{\rm w}^{2/7}$ \citep{cioffi_etal88,thornton_etal98}. Since the highest density in our simulations is about six orders of magnitudes higher than that in \citet{rogers_pittard13}, our simulations require about three orders of magnitude finer spatial resolution than in \citet{rogers_pittard13} to resolve the wind energy feedback, which is computationally prohibitive. We conclude that under the current simulation setup, wind feedback through thermal energy deposition is not appropriate.
Alternatively, wind feedback through momentum deposition can efficiently shut off star formation activities within 2~$\tff$. We find that using volume- or solid angle-weighted schemes for momentum deposition leads to faster cloud disruption than using a mass-weighted scheme. The reason is that the majority of the momentum is deposited to the densest clumps in a mass-weighted scheme and is therefore not able to channel gas out through low-density regions \citep[see also][]{smith_etal18}. In reality, wind momentum should be deposited isotropically around star particles. It is more physically plausible to use volume and solid angle as weights.

\section{Convergence to numerical resolutions}\label{sec:appendix-res}
Hydrodynamic simulations discretize continuous space into  a finite number of resolution elements.
The choice of optimal numerical resolution is to achieve a balance between scientific accuracy and computational costs. To study the convergence of the simulation outcomes to numerical resolutions, we perform GMC simulations of RHO20T runs with $\fb=2$ with different numbers of initial gas elements, $64^3$, $128^3$, and $256^3$, corresponding to target cell masses around $0.191$, $2.38\times10^{-2}$, and $2.98\times10^{-3}\Msun$. The test runs are performed following the same physics as the production runs in the main text, see \autoref{sec:methods}.

We find that the integral star formation efficiencies for the $64^3$, $128^3$, and $256^3$ runs are $\epsint=$~0.612, 0.610, and 0.615, respectively. This consistency suggests that $\epsint$ is not sensitive to mass resolutions but is solely controlled by the force balance between gravitational collapse and gas expulsion by momentum stellar feedback.
As shown in the lower panel of \autoref{fig:sfh-wind}, the star formation histories of the three runs are also in general agreement with each other. The star formation first rises dramatically and peaks at around the free-fall time of the cloud. The only noticeable difference comes from the $64^3$ run. This run shows a narrower star formation history than the other two runs, suggesting delayed star formation at the beginning and an earlier gas removal process after a majority of the stellar mass is formed. The $128^3$ and $256^3$ runs present a more consistent star formation history across the course of GMC evolution.
Therefore, for all production runs present in the main text, we choose $128^3$ as the default number of resolution elements.

\bibliographystyle{mnras}
\bibliography{references} 

\bsp	
\label{lastpage}
\end{document}